\documentclass[preprint]{aastex}
\usepackage{emulateapj5}
\usepackage{apjfonts}
\usepackage{natbib}

\input psfig.tex

\def\aa{{A\&A}}
\def\aas{{ A\&AS}}
\def\aj{{AJ}}
\def\al{$\alpha$}
\def\bet{$\beta$}

\def\annrev{{ARA\&A}}
\def\apj{{ApJ}}
\def\apjs{{ApJS}}
\def\asec{$^{\prime\prime}$}

\def\cc{cm$^{-3}$}

\def\cc{cm$^{-3}$}
\def\e#1{$\times$10$^{#1}$}
\def\etal{{et al. }}

\def\kms{km s$^{-1}$}
\def\lamb{$\lambda$}
\def\lax{{$\mathrel{\hbox{\rlap{\hbox{\lower4pt\hbox{$\sim$}}}\hbox{$<$}}}$}}
\def\gax{{$\mathrel{\hbox{\rlap{\hbox{\lower4pt\hbox{$\sim$}}}\hbox{$>$}}}$}}
\def\simlt{\lower.5ex\hbox{$\; \buildrel < \over \sim \;$}}
\def\simgt{\lower.5ex\hbox{$\; \buildrel > \over \sim \;$}}
\def\lum{erg s$^{-1}$}

\def\mnras{{MNRAS}}

\def\pasp{{PASP}}

\def\percm2{cm$^{-2}$}

\def\solmass{$M_\odot$}

\def\hi{\ion{H}{1}}
\def\hii{\ion{H}{2}}
\def\oiii{[\ion{O}{3}]}

\def\oi{[\ion{O}{1}]}
\def\nii{[\ion{N}{2}]}

\def\sii{[\ion{S}{2}]}

\slugcomment{To appear in {\it The Astrophysical Journal}.}

\shorttitle{PROPERTIES OF EMISSION-LINE NUCLEI}
\shortauthors{HO, FILIPPENKO, \& SARGENT}

\begin{document}
 
\title{A Search for ``Dwarf'' Seyfert Nuclei. VI. Properties of 
Emission-Line Nuclei in Nearby Galaxies}

\author{Luis C. Ho}

\affil{The Observatories of the Carnegie Institution of Washington, 813 Santa
Barbara St., Pasadena, CA 91101}
 
\author{Alexei V. Filippenko}
\affil{Department of Astronomy, University of California, Berkeley, CA
94720-3411}

\and

\author{Wallace L. W. Sargent}
\affil{Palomar Observatory, 105-24 Caltech, Pasadena, CA 91125}
 
\begin{abstract}

We use the database from Paper III to quantify the global and nuclear 
properties of emission-line nuclei in the Palomar spectroscopic survey of 
nearby galaxies.  We show that the host galaxies of Seyferts, LINERs, and 
transition objects share remarkably similar large-scale properties and 
local environments.  The distinguishing traits emerge on nuclear 
scales.  Compared with LINERs, Seyfert nuclei are an order of magnitude more 
luminous and exhibit higher electron densities and internal extinction.   We 
suggest that Seyfert galaxies possess characteristically more gas-rich 
circumnuclear regions, and hence a more abundant fuel reservoir and plausibly 
higher accretion rates.  The differences between the ionization state of the 
narrow emission-line regions of Seyferts and LINERs can be partly explained by 
the differences in their nebular properties.  Transition-type objects are 
consistent with being composite (LINER/\hii) systems.  With very few 
exceptions, the stellar population within the central few hundred parsecs of 
the host galaxies is uniformly old, a finding that presents a serious challenge 
to starburst or post-starburst models for these objects.  Seyferts and LINERs 
have virtually indistinguishable velocity fields as inferred from their line 
widths and line asymmetries.   Transition nuclei tend to have narrower lines 
and more ambiguous evidence for line asymmetries.  All three classes of 
objects obey a strong correlation between line width and line luminosity.  
We argue that the angular momentum content of circumnuclear gas may be an 
important factor in determining whether a nucleus becomes active.  Finally, we 
discuss some possible complications for the unification model of Seyfert 
galaxies posed by our observations.
\end{abstract}


\section{Introduction}

Spectroscopic surveys have shown that emission-line nuclei are very common in 
nearby galaxies (see Ho 1996 and references therein).  Particularly striking 
is the population of galactic nuclei considered ``active,'' by which we mean 
objects whose energy source ultimately derives from nonstellar processes 
associated with accretion onto massive black holes, as is commonly believed 
to be the case for classical Seyfert nuclei and quasars.  According to the 
survey of Ho, Filippenko, \& Sargent (1997b), which is the subject of the 
papers in this series, 43\% of galaxies brighter than $B_T$ = 12.5 mag can be 
considered active galactic nuclei (AGNs) or AGN candidates.  
Among these, the majority (2/3) belong to an enigmatic class of sources 
called low-ionization nuclear emission-line regions (LINERs; Heckman 1980).
Ever since their discovery more than 20 years ago, the physical origin of 
LINERs has been controversial.  While it is now generally acknowledged that a 
significant fraction of LINERs genuinely belong in the AGN family, 
the situation still remains unclear for the class as a whole.  Discussions 
of the competing models and the evidence for and against them can be found 
in the reviews by Filippenko (1996), Ho (1999a, 2002), and Barth (2002).

It should be emphasized that settling the physical origin of LINERs is more 
than of mere phenomenological interest.  Because they are so numerous, 
LINERs could make a tremendous impact on the specification of the faint end 
of the local AGN luminosity function, which is currently very poorly known 
(Huchra \& Burg 1992).  This, in turn, has ramifications for all 
astrophysical issues that depend on the statistics of local AGNs or massive 
black holes.

We completed an extensive  optical spectroscopic survey of the 
central regions of nearly 500 nearby galaxies using the 5-m Hale telescope at 
Palomar Observatory (Filippenko \& Sargent 1985, hereafter Paper~I; 
Ho, Filippenko, \& Sargent 1995, 1997a, 1997b, hereafter Papers~II, III, and V, 
respectively; Ho et al. 1997e, hereafter Paper~IV; Ho, Filippenko, \& Sargent 
1997c, 1997d).  This is the largest and most sensitive survey of its kind 
that has resulted in, among other things, the discovery of an unprecedented 
number of nearby LINERs.  This paper systematically
examines the statistical properties of the sample of AGN candidates in the 
Palomar survey.  The nature of our survey permits a fresh look at 
basic properties that can be either directly measured from our spectra or are
otherwise available from existing sources.  Another objective is to 
compare the traits of LINERs and a related class of nuclei known as 
``transition objects'' (Ho, Filippenko, \& Sargent 1993a; Ho 1996) with those 
of Seyfert nuclei, in order to test the proposition that most LINERs and 
LINER-like sources truly are accretion-powered systems.  We do not consider 
\hii\ (star forming) nuclei at length, except insofar as they illuminate 
the discussion on the AGN candidates; several issues related to \hii\ nuclei 
were treated in Paper~V and Ho et al. (1997c, 1997d).

The quantities analyzed in this paper come from the database in Paper~III.  Ho 
(1996) and Filippenko (1996) have given preliminary discussions of some of the 
results presented here; this paper supersedes the earlier work.  As in 
previous papers in this series, distance-dependent parameters assume a Hubble 
constant of $H_0$ = 75 \kms\ Mpc$^{-1}$.

\section{Analysis}

The ensuing sections give a comparative analysis of a number of global and 
nuclear properties of the various subclasses of AGN candidates.  We wish to 
elucidate the fundamental parameters responsible for the observed diversity 
of spectral characteristics in emission-line nuclei, and ultimately, to gain a 
better physical understanding of nuclear activity in nearby galaxies.  
Following the convention established in the earlier papers of this series, we 
distinguish three categories of AGN-like emission-line nuclei: Seyfert nuclei, 
LINERs, and ``transition objects.''  Seyfert nuclei differ from LINERs 
principally in having higher levels of ionization, and transition objects are 
characterized by spectra that appear to be intermediate between those of 
LINERs and nuclear \hii\ regions.  In optical line-ratio diagnostic diagrams 
such as those introduced by Baldwin, Phillips, \& Terlevich (1981) and 
Veilleux \& Osterbrock (1987), transition objects populate the region between 
traditional \hii\ nuclei and LINERs.  This prompted Ho et al. (1993a) to 
suggest that transition objects are composite systems comprised of a normal 
LINER nucleus whose signal is diluted or contaminated by emission from 
neighboring regions of recent star formation.  If this interpretation is 
correct, then these sources ought to be included in the overall AGN census, to 
the extent that LINERs themselves are genuine AGNs.  We attempt to test both 
of these hypotheses --- that transition nuclei are closely related to 
LINERs and that both of these groups are similar to Seyferts.  In this paper 
we adopt the popular viewpoint that all Seyfert nuclei are bona fide AGNs.

Before proceeding further, we make a few remarks on taxonomy.  Throughout 
this paper, for emphasis, we will cast LINERs, transition objects, and 
Seyferts as distinctly separate groups of emission-line objects.  While this 
approach is useful to highlight general population trends, one should 
recognize that the classification boundaries are fuzzy.  This is obvious from 
inspection of Figure~7 in Paper~III.  LINERs and Seyferts do not form a 
bimodal distribution in ionization; for example, the distribution of the 
\oiii\ \lamb 5007/H\bet\ or \oi\ \lamb 6300/\oiii\ \lamb 5007 ratios is 
continuous for the LINERs and Seyferts in the Palomar survey.  In the same 
vein, the division between LINERs from transition nuclei, or that between 
transition and \hii\ nuclei, is largely arbitrary.  And lastly, contrary to 
popular misconception, not every weak emission-line nucleus in an early-type 
galaxy is a LINER.  As we will later show, in nearby galaxies LINERs are 
typically an order of magnitude underluminous compared to Seyferts, but the 
nuclear luminosities of the two classes overlap generously.

Tables~1{\it a}, 2{\it a}, and 3{\it a}\ summarize basic statistical 
properties for the univariate distributions of various global and nuclear 
parameters.  For each parameter, we give the mean, the standard deviation, and 
the median.  Each subclass of object is listed separately.  For reasons given 
in \S~2.1, our discussion focuses primarily on the spiral galaxies, and in 
particular on a subsample restricted to Hubble types Sab--Sbc ($T = 2-4$); 
however, for completeness, we list also the statistics for the entire sample, 
irrespective of Hubble type.  We evaluate censored data sets (containing upper 
or lower limits) using the Kaplan-Meier product-limit estimator (Feigelson \& 
Nelson 1985).  Two-sample comparisons between subclasses (Tables 1{\it b}, 
2{\it b}, and 3{\it b}) are performed using the Kolmogorov-Smirnov test, and 
as a consistency check, also Gehan's generalized Wilcoxon (hereafter Gehan) 
test (Isobe, Feigelson, \& Nelson 1986).  We quote the significance in terms 
of the probability of rejecting the null hypothesis that the two distributions 
are drawn from the same parent population, $P_{\rm null}$.   We consider two 
distributions to be statistically different if $P_{\rm null} < 5\%$.   The 
significance of the difference between two means is evaluated using Student's 
$t$ test (Press et al. 1986).  For these three tests, we designate the 
probability of rejecting the null hypothesis as $P_{K}$, $P_G$, and $P_t$, 
respectively.

\subsection{Host Galaxy Properties}

The host galaxies of the AGN candidates show a surprising degree of 
homogeneity in their large-scale, global properties.  We found in Paper~V that 
all three subclasses inhabit galaxies of very similar morphologies, mostly 
Hubble types ranging from ellipticals to lenticulars and early-type, 
bulge-dominated spirals (Sbc and earlier).  The only noticeable difference is 
that a larger fraction of elliptical and S0 galaxies contains pure LINERs, 
and transition objects tend to be found in galaxies of slightly later Hubble 
type.  Table 1{\it a}\ shows that the mean and median morphological index 
($T$) of LINERs is about two units earlier than those of transition objects 
and Seyfert galaxies.  When we restrict the comparison to the spiral 
subsample, the $T$ distributions for LINERs and Seyferts become quite similar, 
but transition objects remain statistically different compared to LINERs 
(Table 1{\it b}), by $\Delta T \approx 1$.  Since many global as well as 
nuclear properties vary systematically with Hubble type, we must be wary of 
potential selection effects that can be introduced when comparing samples 
mismatched in Hubble type.  To mitigate this problem, we will concentrate our 
analysis on a highly restricted subsample comprised only of galaxies with 
Hubble types Sab--Sbc ($T = 2-4$; hereinafter referred to as ``the Sb 
subsample'')\footnote{We are forced to violate this rule when examining issues 
related to type 1 versus type 2 AGNs, objects with and without detectable 
broad emission lines, respectively.  There would otherwise be insufficient 
objects for a meaningful analysis.  We use, instead, the full subsample of 
spirals, which fortunately have statistically similar distributions of Hubble 
types for each of the two types of LINERs and Seyferts (see Table 1{\it b}).  
The subsample of spirals contains 11 LINER 1s, 39 LINER 2s, 18 Seyfert 1s, and 
20 Seyfert 2s.}.  This range was chosen as a compromise between the desire to 
isolate a morphologically homogeneous sample for each AGN subclass and the 
need to retain sufficient numbers for meaningful statistical analysis.  With 
this choice, all three AGN subclasses have a mean and median $T \approx 3$.   
The Sb subsample contains 23 LINERs, 32 transition objects, and 19 Seyferts.  
For comparison, we also consider a matched subsample of 36 \hii\ nuclei, 
constructed by combining all the $T = 2-3$ objects in the parent sample with a 
randomly chosen subset of $\sim 1/3$ of the $T = 4$ objects; we did not use 
the full sample of \hii\ nuclei with $T = 4$ hosts because this morphological 
type contains a higher percentage of \hii\ nuclei than AGNs.

The similarity among the three AGN subclasses can be further discerned in their 
absolute optical luminosities, which are typically close to, or somewhat in 
excess of, $L^*$.  As shown in Table~1{\it b}, they also have statistically 
similar distributions of bulge luminosities [$M_B$(bul)], neutral hydrogen 
content (\hi\ mass normalized to the optical luminosity, $M_{\rm HI}/L^0_B$), 
as well 

\begin{figure*}[t]
\centerline{\psfig{file=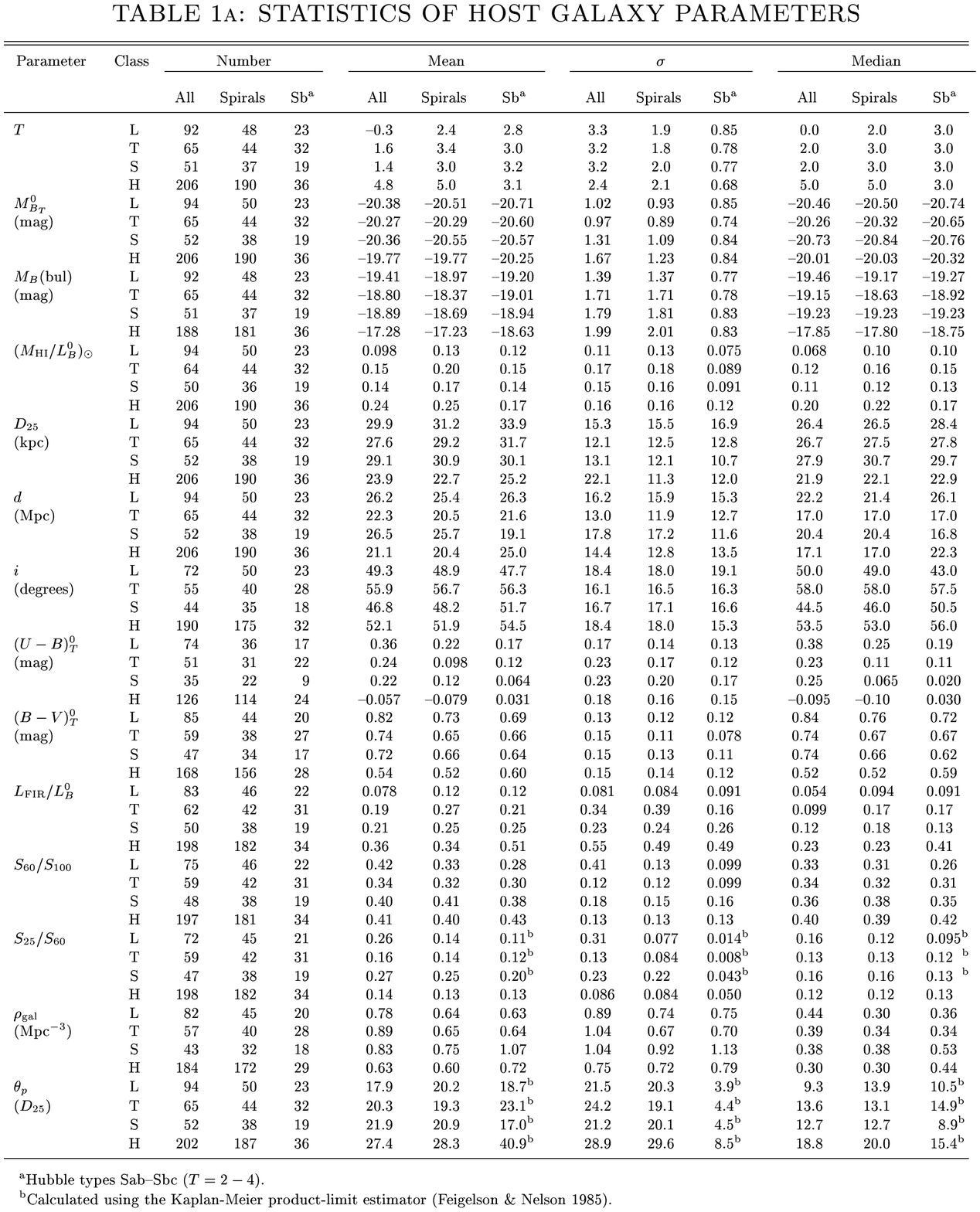,width=18.5cm,angle=0}}
\end{figure*}
\clearpage

\begin{figure*}[t]
\centerline{\psfig{file=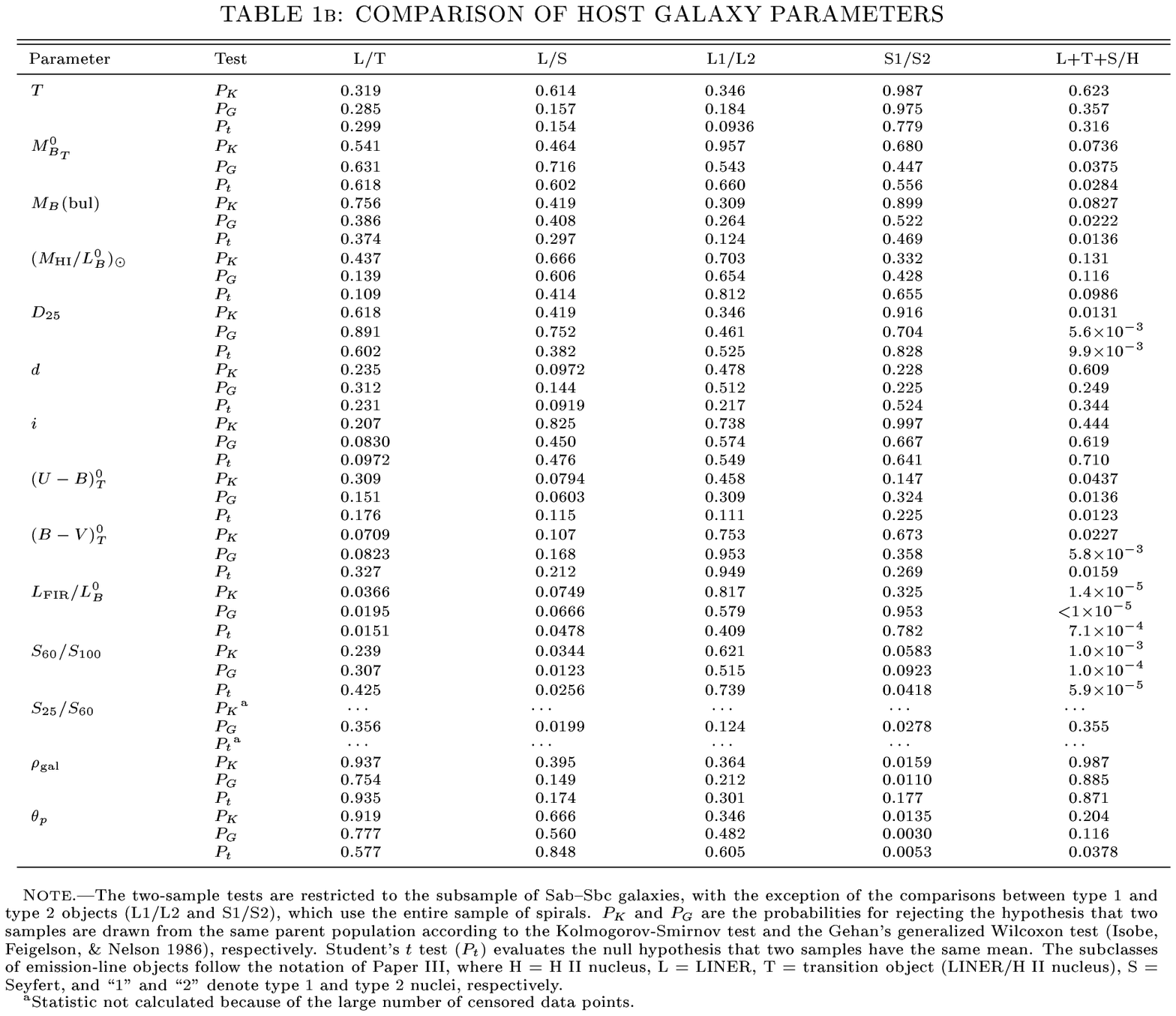,width=18.5cm,angle=0}}
\end{figure*}
\clearpage

\vskip 0.3cm

\psfig{file=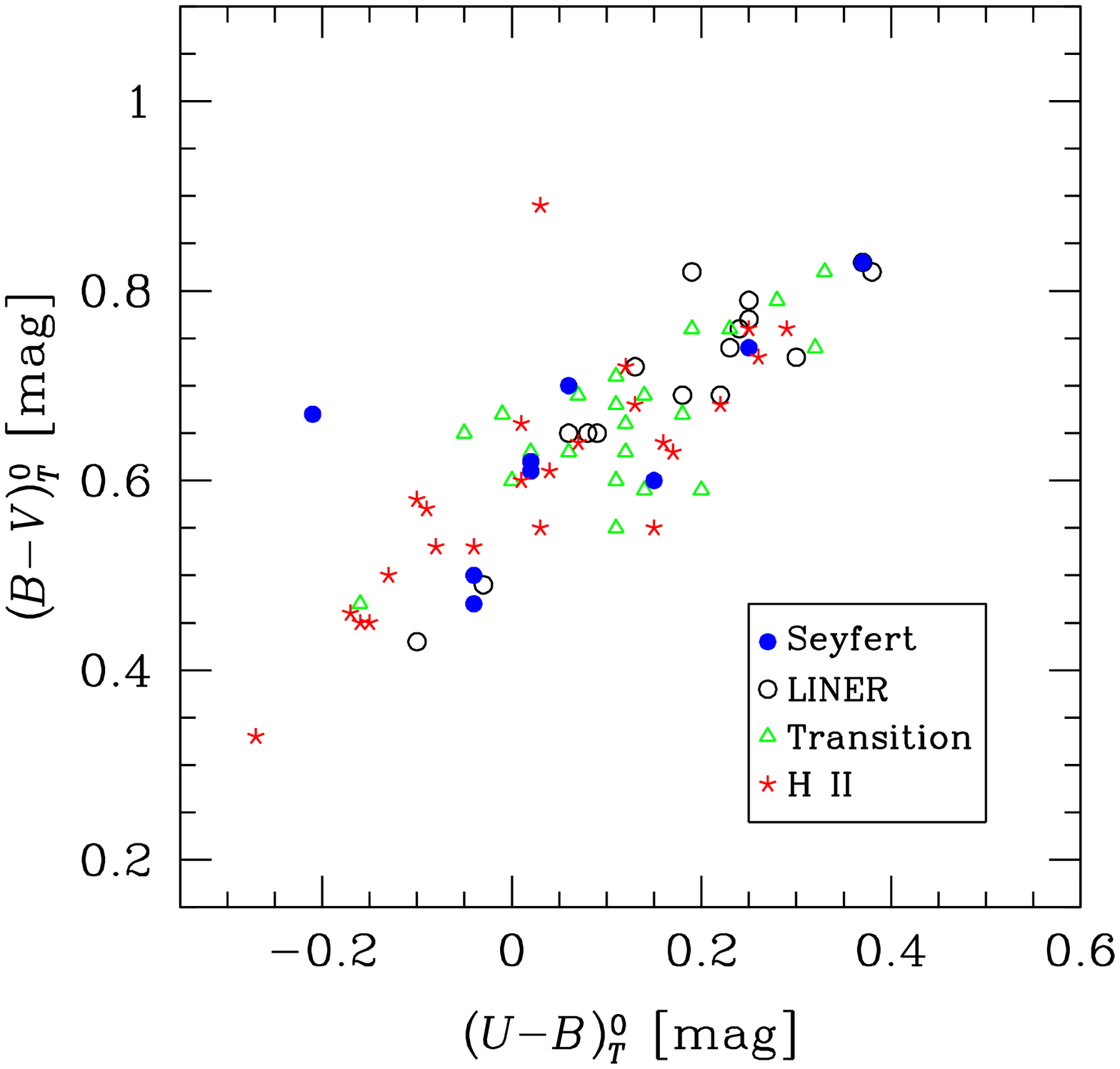,width=8.5cm,angle=0}
\figcaption[fig1.ps]{
Optical colors of Seyferts (solid circles), LINERs (open circles), transition
objects (triangles), and H~II nuclei (stars).  The colors are based on
integrated magnitudes corrected to face-on orientation.  Only Sab--Sbc
galaxies are shown.
\label{fig1}}
\vskip 0.3cm

\noindent
as optical isophotal diameters ($D_{25}$).  Transition objects may be 
marginally more highly inclined than LINERs, although the statistical 
significance is low ($P_{K}$ = 21\%, $P_G$ = 8.3\%, and $P_t$ = 9.7\%), 
whereas LINERs are indistinguishable from Seyfert galaxies; we will revisit
this point in \S~3.3.  

The most interesting differences emerge from inspection of the broad-band, 
integrated optical colors (Fig.~1) and the far-infrared (FIR) properties 
(Fig.~2).  Two patterns deserve attention.  Relative to LINERs, transition 
objects exhibit enhanced FIR emission (for a given optical luminosity), at the 
level of $P_{\rm null} \approx$ 1\%--4\%.  There are no obvious differences in 
their FIR colors.  This might reflect an elevated level of star formation, 
either global or nuclear, in transition objects.  Consistent with this 
hypothesis, transition objects show mildly bluer integrated $U-B$ and $B-V$ 
colors, although the level of significance is low.  Interestingly, Seyfert 
galaxies too tend to have higher normalized FIR luminosities and somewhat 
bluer optical colors compared to LINERs.  But these two groups also occupy 
slightly different loci in the FIR color-color diagram (Fig.~2{\it a}): 
Seyferts have statistically ``hotter'' FIR colors  (higher $S_{60}/S_{100}$ 
and $S_{25}/S_{60}$ flux density ratios).  Although one might also attribute 
these characteristics to enhanced star formation in the Seyfert population, we 
note that the $S_{25}/S_{60}$ ratios of the Seyferts are generally larger than 
those in \hii\ nuclei. This suggests that the enhanced FIR emission and hotter 
FIR colors in Seyferts may instead be due to a higher level of AGN activity 
in these objects.

We note that the two types of Seyfert galaxies in our sample
differ in their FIR properties: Seyfert 1s possess higher $S_{60}/S_{100}$
and $S_{25}/S_{60}$ ratios than Seyfert 2s.  We will discuss the implications
of this result in \S~3.5.  The two Seyfert 
types show no other obvious differences in global properties.  LINER 1s and 
LINER 2s cannot be distinguished on the basis of their global properties.

\subsection{Environment}

Tidal interaction with neighboring galaxies is often cited as a possible 
mechanism for triggering nuclear activity.  Paper III gives two parameters 
that can be used to evaluate the importance of this effect in our sample: 
(1) $\rho_{\rm gal}$, defined by Tully (1988) as the density of all galaxies 
brighter than $M_B$ = --16 mag in the object's local vicinity, and 
(2) $\theta_p$, the projected angular separation to the nearest neighbor 
within a magnitude difference of $\pm$1.5 mag and a velocity difference of 
$\pm$500 \kms, measured in units of the isophotal angular diameter of
the primary galaxy, $D_{25}$.  After excluding the elliptical and S0 
galaxies, whose overrepresentation among LINERs might bias that sample 
because of the morphology-density relation (Dressler 1980; Postman \& Geller 
1984), we find that the local environment, as measured by $\rho_{\rm gal}$ 
and $\theta_p$, has no impact on the spectral classification of the nucleus.  
The same conclusion holds for the Sb subsample.  

The only exception concerns the two subtypes of Seyfert galaxies: to a 
relatively high level of significance, Seyfert 1s inhabit denser environments 
than Seyfert 2s.  The mean and median galaxy density for Seyfert 1s is 1 
and 0.6 Mpc$^{-3}$, compared with 0.5 and 0.3 Mpc$^{-3}$ for Seyfert 2s.
Even more striking is the projected distance to the nearest neighbor. 
For Seyfert 1s, the mean and median value of $\theta_p$ is $\sim$12 and 7 
$D_{25}$; for Seyfert 2s, the corresponding values are $\theta_p \approx$ 
29 and 23 $D_{25}$.  In terms of $\theta_p$, the differences between the two 
samples are significant at the level of $P_{K}$ = 1.3\%, $P_G$ = 0.3\%, and 
$P_t$ = 0.5\%.

\subsection{Nuclear Properties}

We consider several nebular parameters that might provide clues to the 
physical conditions of the line-emitting regions (see Tables 2{\it a}\ and 
2{\it b}).  We find no significant differences between LINERs and transition 
objects in terms of line luminosity\footnote{Some of the objects in the
Palomar survey were observed under nonphotometric conditions.  Whenever
possible, we supplemented the database in Paper~III with H\al\ luminosities
published in the literature.  A list of these data is given in the Appendix.}
($L_{{\rm H}\alpha}$; Fig.~3), line equivalent width [EW(H\al)], or electron
density ($n_e$; Fig.~4).  An interesting trend seen in the sample of 
transition objects, but absent from the others, is the tendency for more 
highly inclined sources to show larger amounts of internal reddening 
(Fig.~5).  The Kendall's $\tau$ correlation coefficient between 
$E(B-V)_{\rm int}$ and $i$ is $r = 0.70$, significant at the level of 99\%.  
The source of dust opacity in transition objects evidently is somehow coupled 
to the large-scale disk component of the galaxy.  On average, transition 
objects tend to be marginally more reddened than LINERs 
[$\langle E(B-V)_{\rm int}\rangle$ = 0.36 vs. 0.19 mag; $P_t$ = 4.0\%]; the
reddening distributions for the two classes may be inconsistent with being
drawn from the same parent population ($P_{K}$ = 3.6\% and $P_G$ = 9.7\%;
Fig.~6).

By contrast, LINERs stand out from Seyfert nuclei in several important
respects.  LINERs have weaker emission lines [$\langle$EW(H\al)$\rangle$
smaller by a factor of $\sim$8], lower H\al\ luminosity (factor of $\sim$11 in
$\langle L_{{\rm H}\alpha}\rangle$), lower density (factor of $\sim$3 in
$\langle n_e \rangle$), and lower internal reddening [factor of $\sim$2 in
$\langle E(B-V)_{\rm int}\rangle$].   The differences between the two samples
for all these parameters are highly significant according to the
Kolmogorov-Smirnov, Gehan, and Student's $t$ tests when the full sample
of spirals is considered.  The statistical significances are somewhat
diminished for the Sb subsample, probably because of the more limited
statistics.  Nevertheless, even a visual inspection of Figures 3, 4, and 6
leaves little doubt that LINERs and Seyferts have systematically different
nebular properties.  As discussed

\begin{figure*}[t]
\centerline{\psfig{file=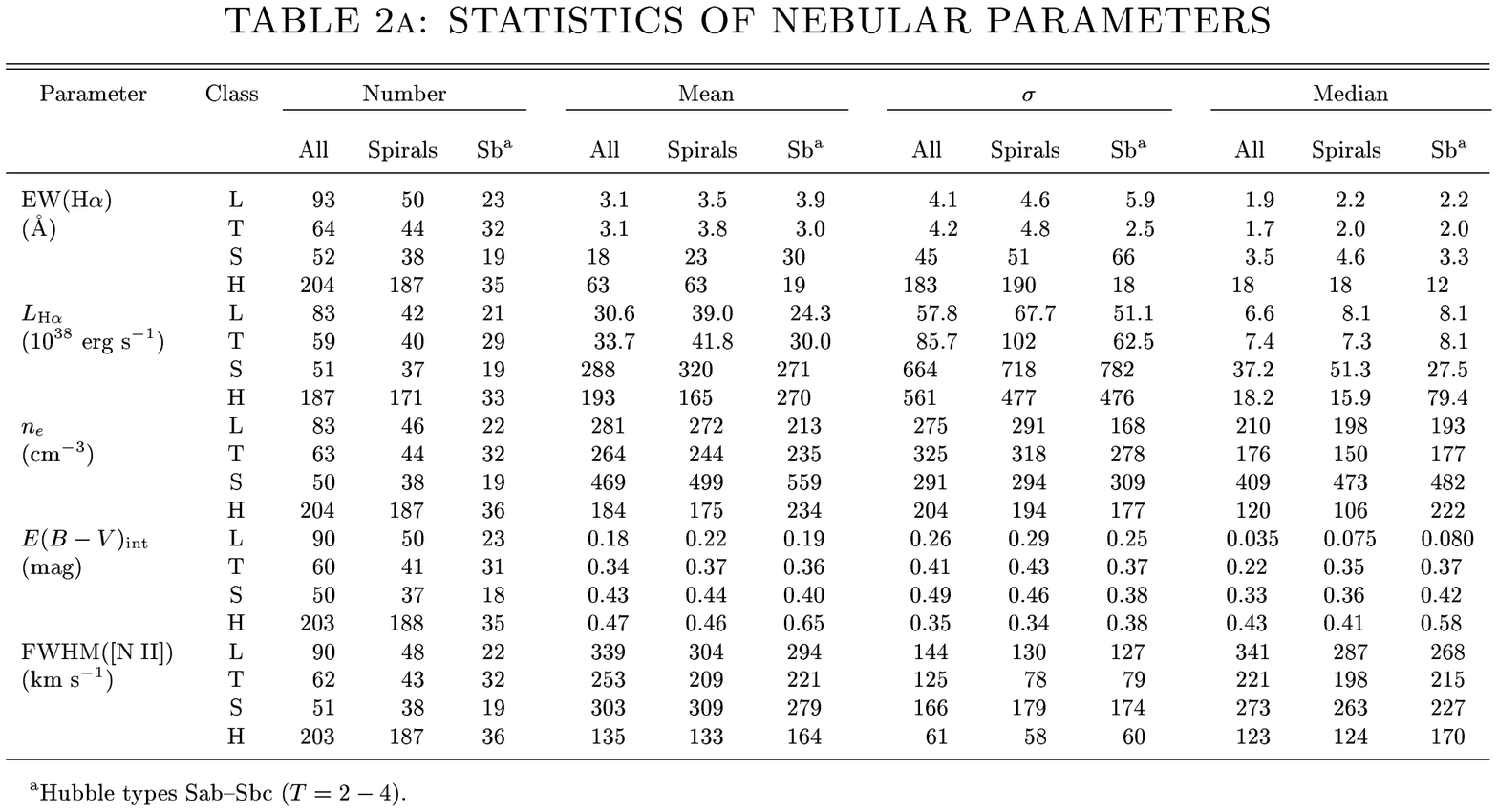,width=18.5cm,angle=0}}
\end{figure*}

\begin{figure*}[t]
\centerline{\psfig{file=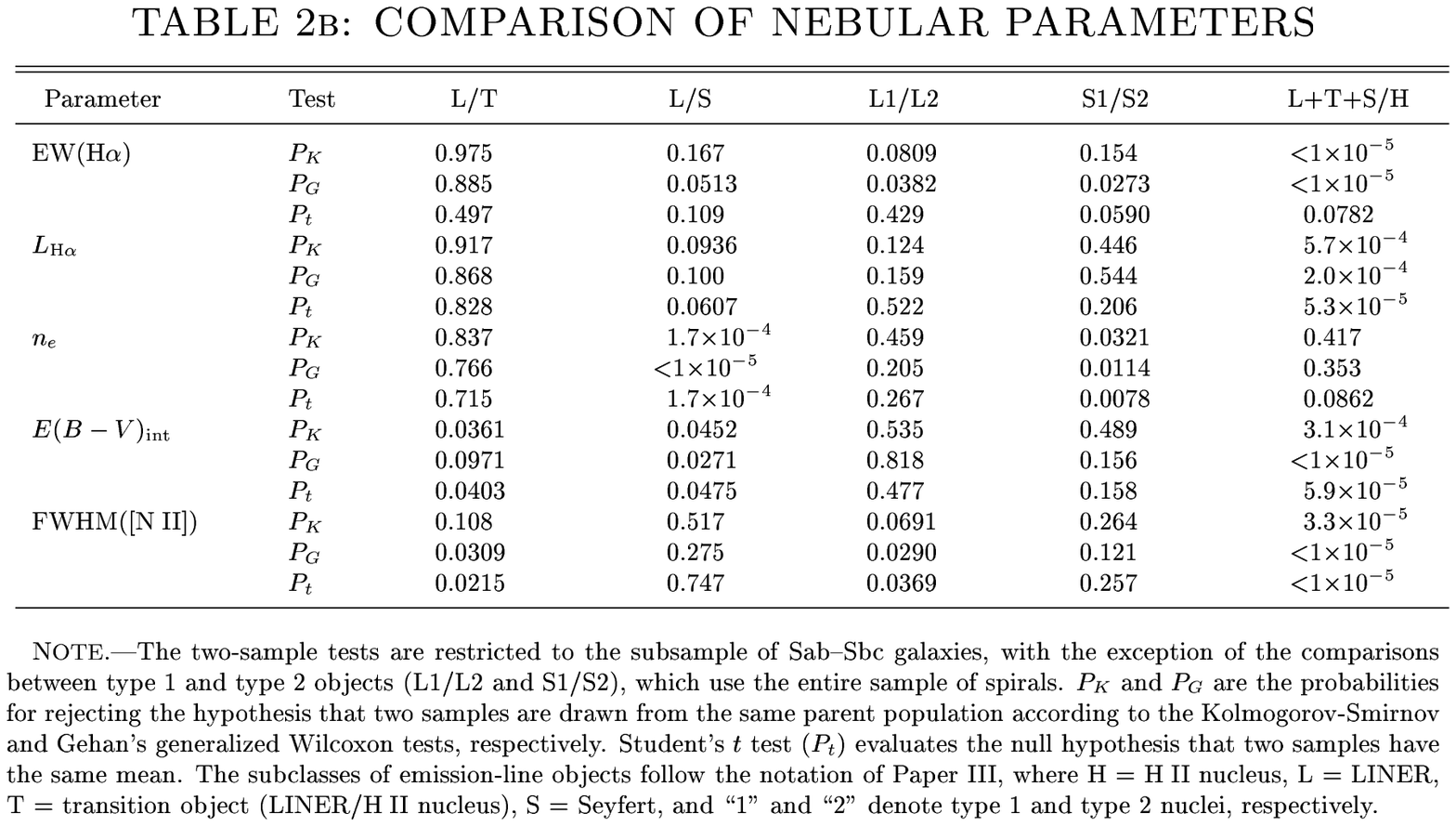,width=18.5cm,angle=0}}
\end{figure*}
\vskip 0.3cm

\noindent
in more detail in \S~3.2, these trends 
strongly suggest that the host galaxies of LINERs, and in particular their 
circumnuclear regions, contain less gaseous material than the host galaxies of 
Seyfert nuclei.  We will argue that this difference may translate into a 
difference in the amount of fuel available to power the nuclei.

We note that the electron densities given here, which were derived from the 
ratio of the \sii\ \lamb\lamb 6716, 6731 lines using the code of Shaw \& 
Dufour (1993) and the S$^+$ atomic data of Cai \& Pradhan (1993), are 
significantly lower than the values typically quoted in older studies of 
LINERs (e.g., Stauffer 1982b; Keel 1983c; Phillips et al.  1986) and Seyferts 
(e.g., Koski 1978).  Most of the differences can be attributed to revisions 
in the atomic data.

\subsection{Kinematic Properties of the Nuclear Gas}

The kinematic information contained in the profiles of the emission lines 
provides additional constraints on the physical conditions of the 
circumnuclear environment.  Previous 

\vskip 0.3cm

\begin{figure*}[t]
\centerline{\psfig{file=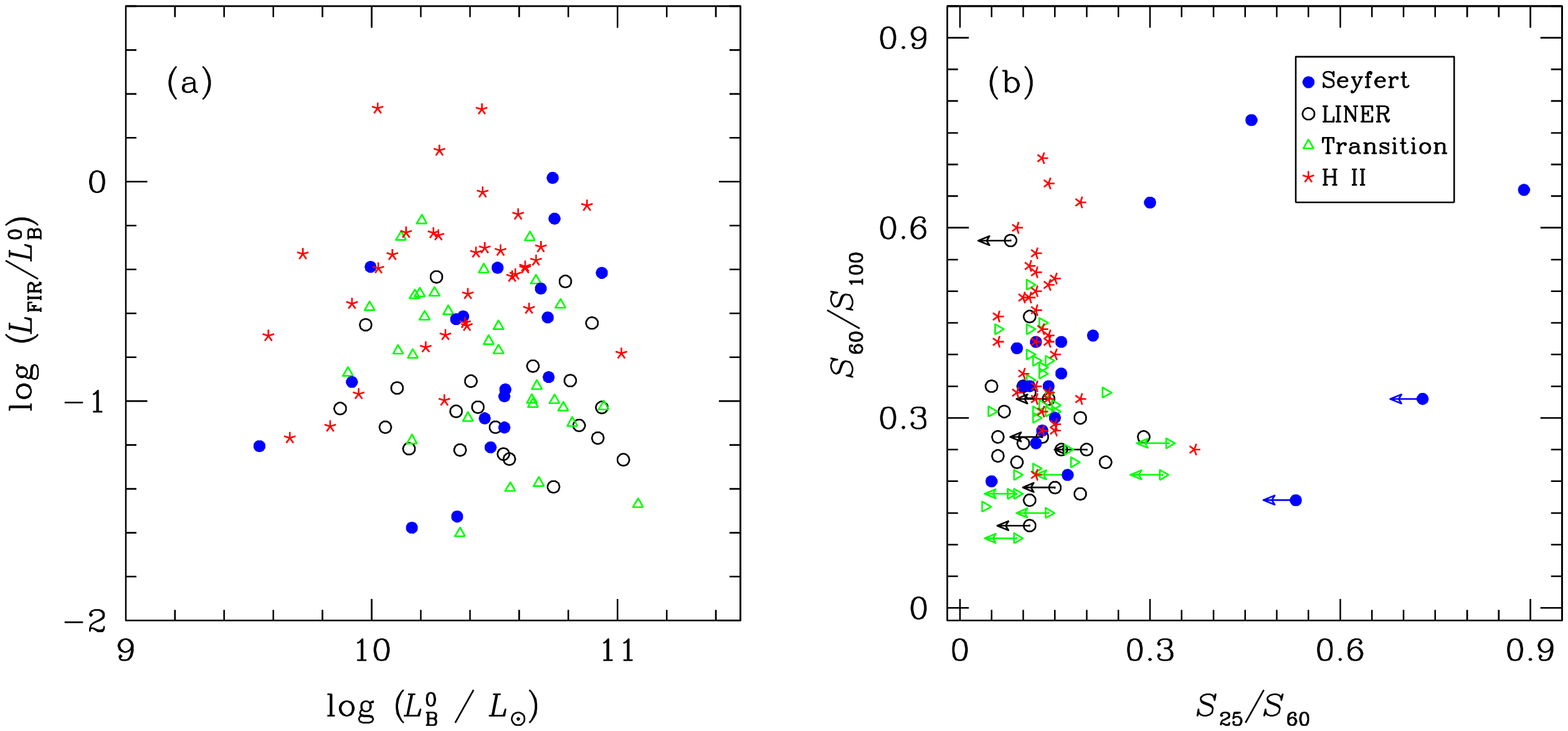,width=19.5cm,angle=270}}
\figcaption[fig2.ps]{
FIR properties of Seyferts (solid circles), LINERs (open circles), transition
objects (triangles), and \hii\ nuclei (stars) derived from {\it IRAS}\
data.  Only Sab--Sbc galaxies are shown.  Panel ({\it a}) plots the FIR
luminosity, $L_{\rm FIR}$, normalized to the inclination-corrected $B$-band
luminosity, $L_B^0$, versus $L_B^0$.  Panel ({\it b}) plots the FIR colors
$S_{60}/S_{100}$ versus $S_{25}/S_{60}$.  Note that Seyferts have
systematically stronger FIR emission and hotter FIR colors than LINERs.
\label{fig2}}
\end{figure*}
\vskip 0.3cm

\noindent
kinematic studies have concentrated 
almost exclusively on Seyferts.  Aside from a small handful of relatively 
crude line width measurements (e.g., Dahari \& De Robertis 1988; Whittle 1993, 
and references therein), little else is known about the line profiles of 
LINERs as a class.  Scarcer still are data for transition objects.  Indeed, 
with a few exceptions, the full width at half maximum (FWHM) of the forbidden 
lines in LINERs 

\vskip 0.3cm

\psfig{file=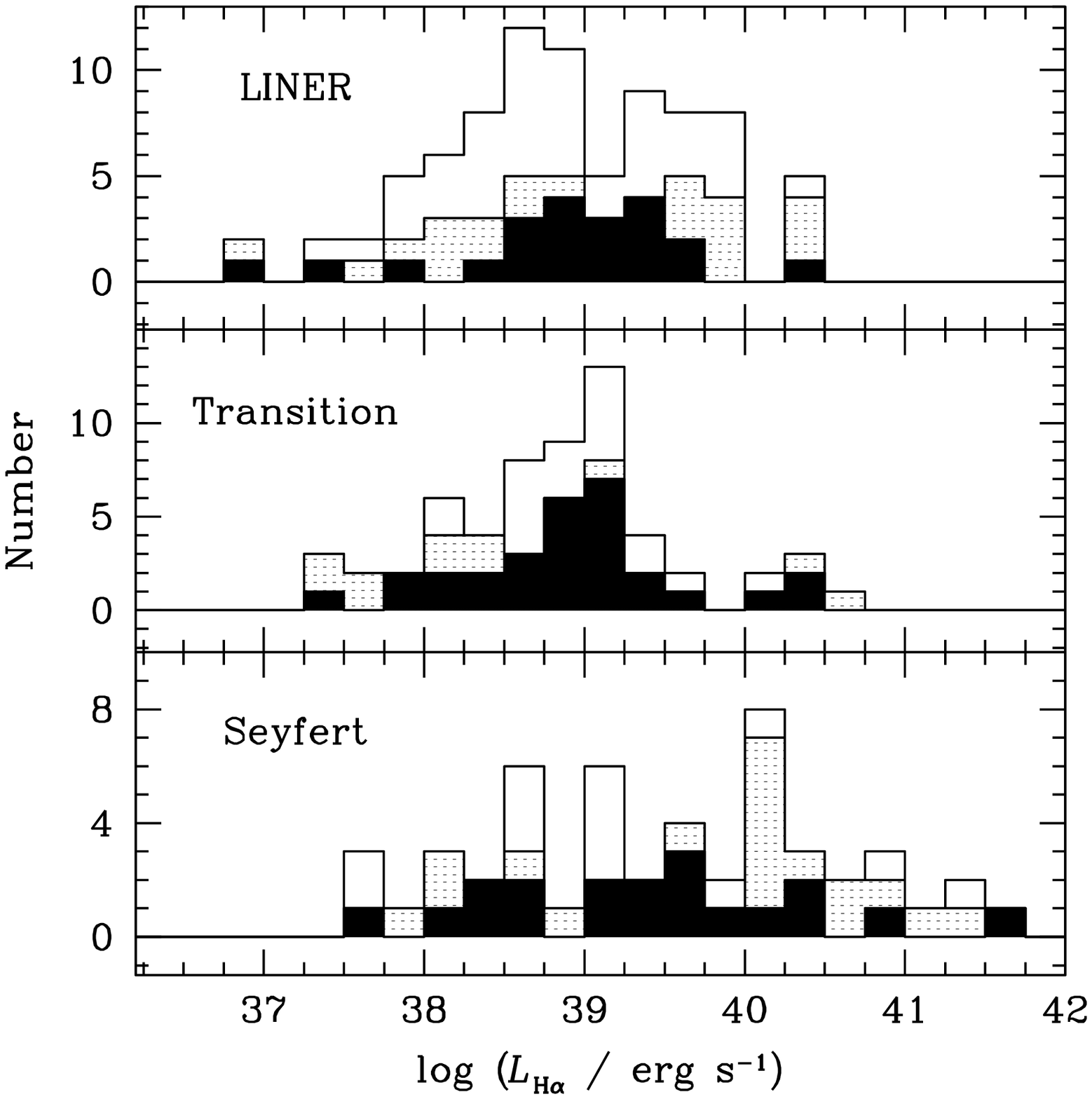,width=8.5cm,angle=0}
\figcaption[fig3.ps]{
Distribution of extinction-corrected luminosities of the narrow H\al\ emission
line for the different classes of AGNs.  The bins are separated by 0.25 in
logarithmic units.  The open histograms plot the E and S0 galaxies, the shaded
histograms plot the spiral galaxies excluding Sab--Sbc, and the solid
histograms plot the Sab--Sbc galaxies.
\label{fig3}}
\vskip 0.3cm

\noindent
rarely exceeds 500 \kms, the typical resolution of many 
previous surveys.  Since the initial study of Heckman (1980), it has commonly 
been assumed that the line widths of LINERs are roughly comparable to those of 
Seyferts (Wilson \& Heckman 1985; Whittle 1985b, 1993), although Stauffer 
(1982b) has remarked, admittedly based on very small number statistics, that 

\vskip 0.3cm

\psfig{file=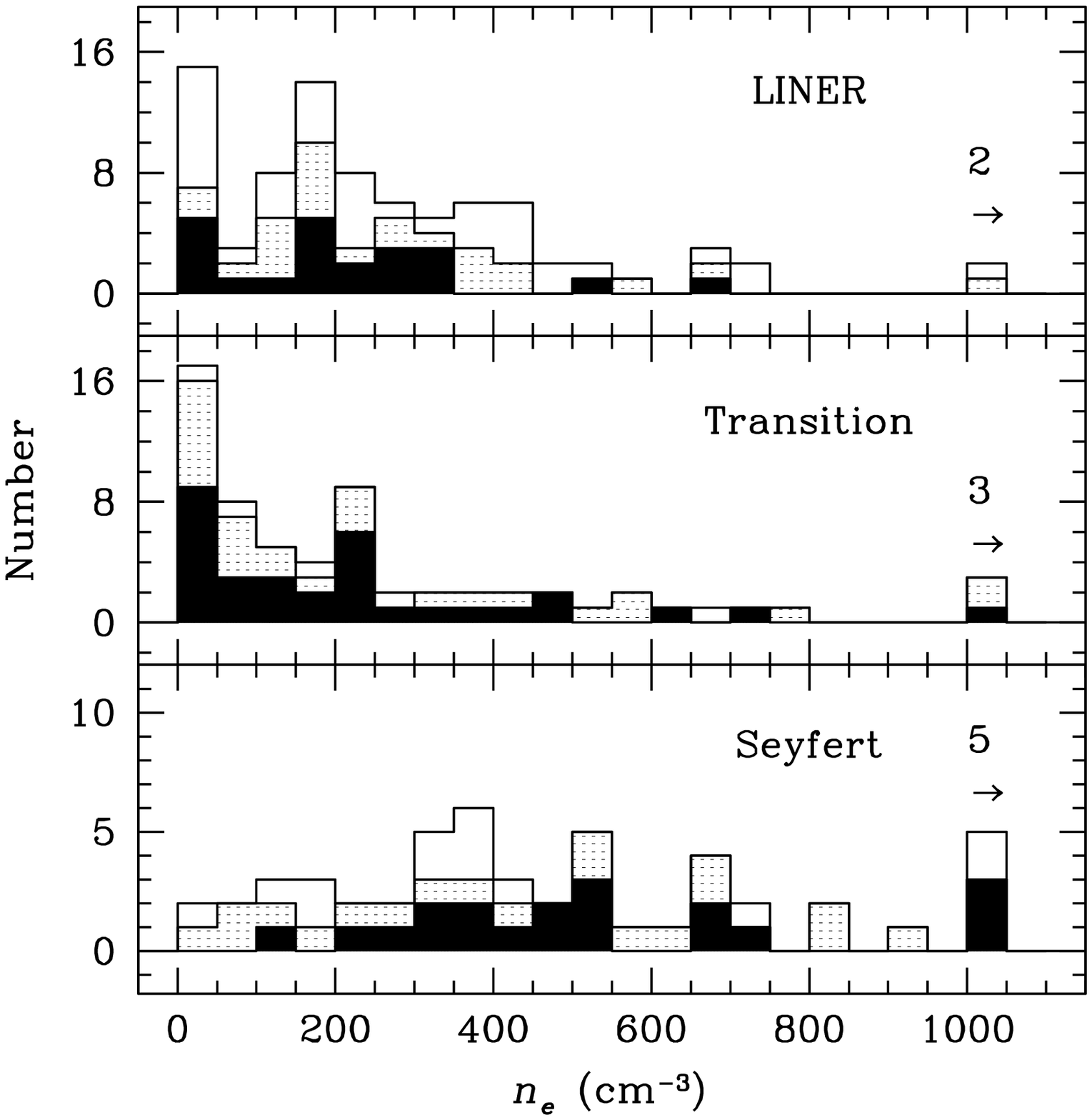,width=8.5cm,angle=0}
\figcaption[fig4.ps]{
Distribution of electron densities, derived from [S~II] \lamb\lamb 6716, 6731,
for the different classes of AGNs.  The bins are separated by units of 50 \cc.
The last bin contains all objects with $n_e$ $>$ 1000 \cc, and the number of
such objects is indicated.  The open histograms plot the E and S0 galaxies,
the shaded histograms plot the spiral galaxies excluding Sab--Sbc, and the
solid histograms plot the Sab--Sbc galaxies.
\label{fig4}}
\vskip 0.3cm

\vskip 0.3cm

\psfig{file=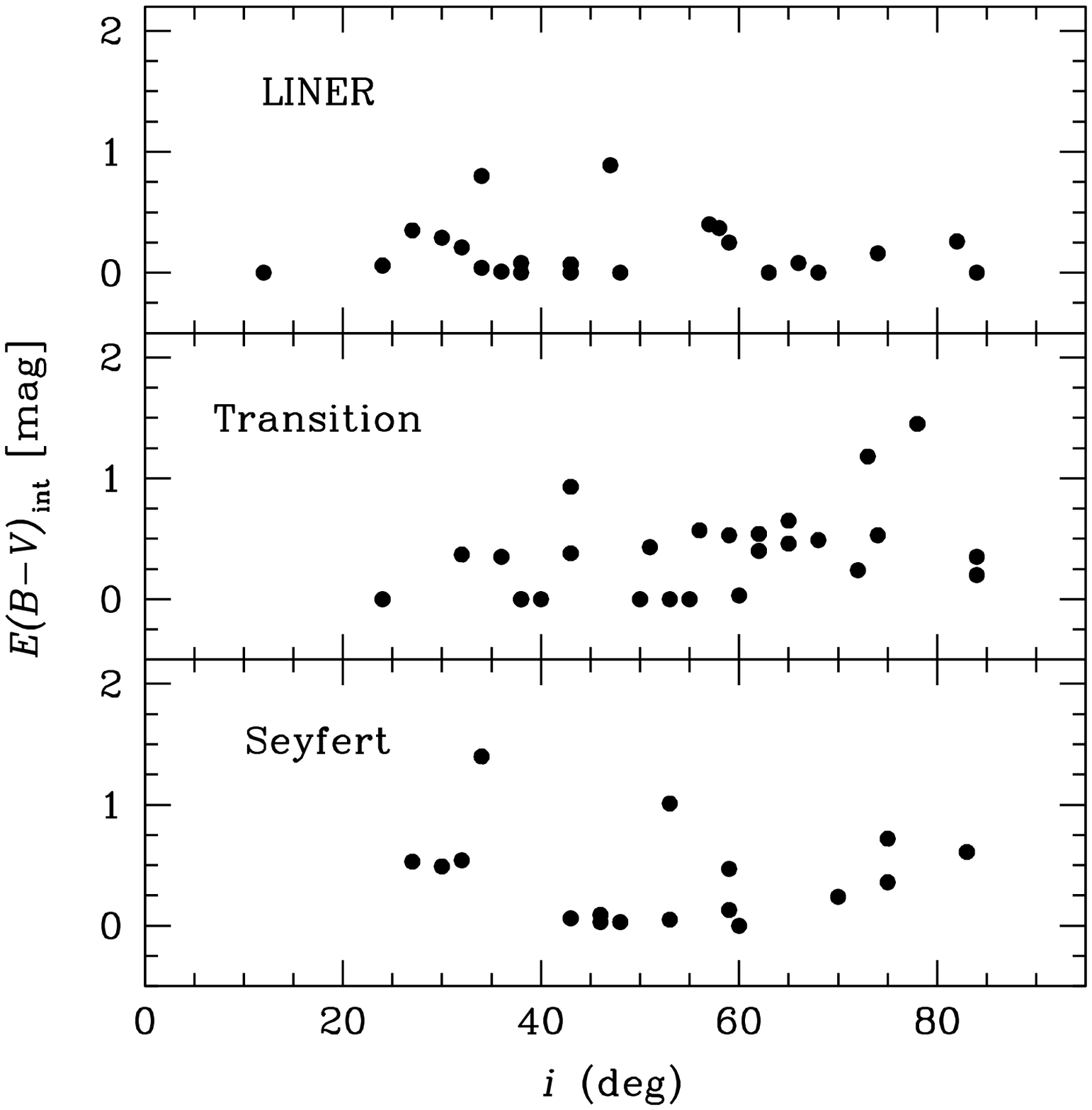,width=8.5cm,angle=0}
\figcaption[fig5.ps]{
Internal reddening of the line-emitting regions plotted against the
inclination angle of the host galaxies.  The inclination angles, tabulated
in Paper~III, are derived from the isophotal axial ratios.  A slight trend
toward larger reddening values in high-inclination (edge-on) systems may be
present in transition objects, while no correlation is apparent for LINERs
and Seyferts.  Only Sab--Sbc galaxies are shown.
\label{fig5}}
\vskip 0.3cm

\noindent
LINERs seem to have broader lines than Seyferts.  The study of Phillips et al. 
(1986), whose spectral resolution is comparable to that of the Palomar survey, 
also has clearly shown that the typical line widths in LINERs are 
substantially smaller than what Heckman had first thought.

Two simple kinematic parameters can be extracted from the narrow emission 
lines: their width and sense of asymmetry.  As explained in Paper~III, in the 
Palomar survey the line of choice for profile measurement is \nii\ \lamb 6583, 
whose width is represented by its FWHM.  

\subsubsection{Line Widths}

The line widths range from being nearly unresolved ($\sim$100 \kms) to 
500--700 \kms, with an average FWHM of 339, 253, and 303 \kms, respectively, 
for the entire sample of LINERs, transition objects, and Seyferts (Fig.~7).   
After isolating the Sb subsample, the values become 294, 221, and 279 \kms, 
respectively.  Transition objects tend to have narrower lines than LINERs, 
with marginal significance ($P_{K}$ = 11\%, $P_G$ = 3.1\%, and $P_t$ = 
2.2\%).  LINERs and Seyferts, on the other hand, are 
virtually identical in terms of their line widths. 

In the central regions of galactic bulges, the velocity dispersion of the 
ionized gas generally traces the velocity dispersion of the stars (Bertola 
et al. 1984; Whittle 1992; Nelson \& Whittle 1996).  Thus, at first 
sight, the narrower lines seen in transition objects appear to indicate that 
the bulges of their host galaxies have systematically shallower gravitational 
potential wells.  This interpretation, however, conflicts with our knowledge 
of the host galaxies (\S~2.1) ---  the morphological types, absolute 
magnitudes, and especially the bulge luminosities are very similar among the 
three AGN subclasses.   We suggest another explanation: the ionized gas in 
transition objects is kinematically colder than in LINERs or Seyferts.  This 
may arise, for example, if the line-emitting clouds preferentially lie in a 
rotationally supported disk.

\vskip 0.3cm

\psfig{file=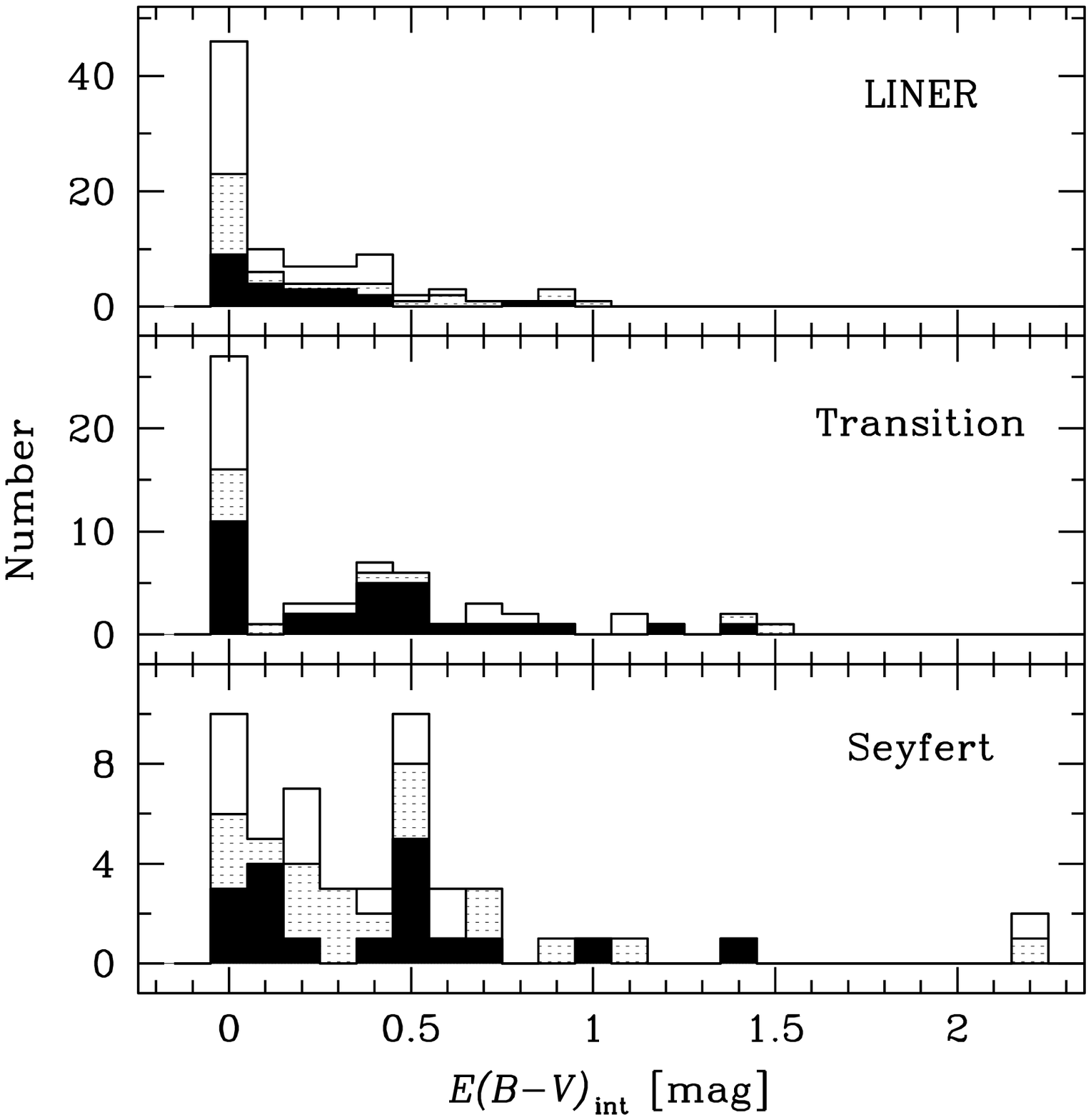,width=8.5cm,angle=0}
\figcaption[fig6.ps]{
Distribution of internal reddening values, inferred from comparison of
the observed H\al/H\bet\ ratios to the theoretical value of 3.1 (Halpern
\& Steiner 1983) and assuming the Galactic extinction curve of Cardelli,
Clayton, \& Mathis (1989).  The bins are separated by units of 0.1 mag.
The open histograms plot the E and S0 galaxies, the shaded
histograms plot the spiral galaxies excluding Sab--Sbc, and the solid
histograms plot the Sab--Sbc galaxies.
\label{fig6}}
\vskip 0.3cm

As first noticed by Phillips, Charles, \& Baldwin (1983), and later quantified 
more extensively by Whittle (1985b, 1992), the luminosities of the forbidden 
lines in Seyfert nuclei positively correlate with their widths.  The 
interpretation of this correlation has been controversial, however, mainly 
because of the existence of other mutual dependences between line width, line 
luminosity, and radio power (Wilson \& Heckman 1985).  Whittle (1992) suggests 
that the fundamental parameter driving 
all these correlations is the bulge mass (or central gravitational potential) 
of the host galaxy.  The Seyferts in our sample also display a strong 
correlation between line luminosity and line width (Fig.~8), extending it down
in luminosity by over two orders of magnitude compared to previously published 
samples.   For the \oiii\ \lamb 5007 line, Whittle (1985b) reported a very 
steep relation with a slope of $\sim 6.4$.  We fitted our data using an 
unweighted linear regression line, calculated using the ordinary least-squares 
solution bisector with jackknife resampling (Feigelson \& Babu 1992), of the 
form $\log L_{{\rm H}\alpha}\,=\,a \log {\rm FWHM([N~II])}\,+\,b$.  We find 
($a,\,b$) = ($4.7 \pm 0.6,\,28.2 \pm 1.4$) for the spiral subsample and 
($a,\,b$) = ($5.7 \pm 0.9,\,25.6 \pm 2.3$) for the entire sample.  Figure 8 
shows --- to our knowledge for the first time --- that LINERs also obey the 
line luminosity-width correlation; Wilson \& Heckman (1985) previously 
concluded that they do not.  Interestingly, the slope of the correlation in 
LINERs is noticeably shallower than in Seyferts; the fits for the spiral and 
entire sample yield, respectively, ($a,\,b$) = ($3.2 \pm 0.8,\,31.3 \pm 2.0$) 
and ($2.7 \pm 0.5,\,32.3 \pm 1.3$).   Transition objects behave essentially 
the same as the LINERs: ($a,\,b$) = ($3.5 \pm 0.5,\,30.7 \pm 1.2$) for the 
spirals and ($a,\,b$) = ($2.9 \pm 0.6,\,32.0 \pm 1.4$) for the entire sample.
Finally, combining all three AGN subtypes, we obtain ($a,\,b$) = 
($4.0 \pm 0.3,\,29.5 \pm 0.8$) and ($3.6 \pm 0.4,\,30.4 \pm 1.0$) for the 
spiral and entire sample, respectively.  (We confirmed that the Sb subsample 
gives very similar fits as the full spiral sample.)

\vskip 0.3cm

\psfig{file=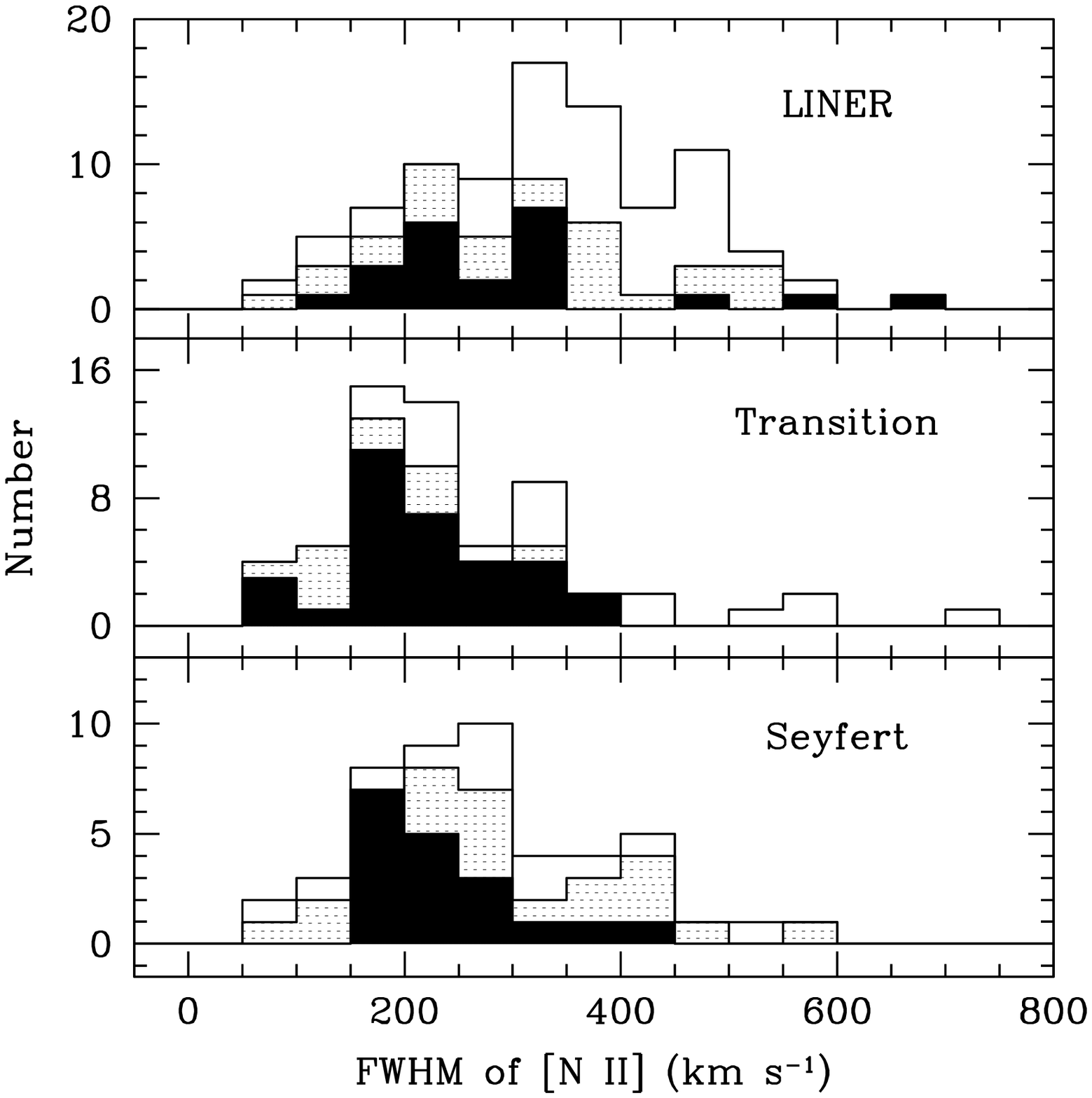,width=8.5cm,angle=0}
\figcaption[fig7.ps]{
Distribution of line widths (FWHM) of the [N~II] \lamb 6583 line for the
different classes of AGNs.  The line widths have been corrected for
instrumental resolution.  The bins are separated by units of 50 \kms.
The open histograms plot the E and S0 galaxies, the shaded
histograms plot the spiral galaxies excluding Sab--Sbc, and the solid
histograms plot the Sab--Sbc galaxies.
\label{fig7}}
\vskip 0.3cm

Keel (1983c) found that in his sample the widths of the forbidden lines are
well correlated with galaxy inclination angle, which suggests that motion in 
the plane of the galaxy disk dominates the velocity field of the narrow-line 
region (NLR).  The present data set does not support this conclusion; no 
dependence between FWHM(\nii) and galaxy inclination angle is seen for the 
AGN subclasses, either individually or combined.  Other studies have come to 
a similar conclusion (Heckman et al. 1981; Wilson \& Heckman 1985; Whittle
1985b; V\'eron \& V\'eron-Cetty 1986).  One can infer that
either the NLR does not have a disklike geometry in the plane of the
galactic disk, or that the component of the velocity field in the
galactic plane contributes only a portion of the total observed line widths
(Whittle 1985b, 1992).

We mention, in passing, that in light of the dependence of line width on 
luminosity, it is hardly surprising that the typical Seyfert nucleus has much 
narrower lines than conventionally assumed.  Hence, the criterion for 
distinguishing Seyfert 2 nuclei from ``normal'' emission-line nuclei (i.e., 
\hii\ nuclei) based on the widths of the narrow lines, either as originally 
proposed by Weedman (1970, 1977), or as later modified by Balzano \& Weedman 
(1981), Shuder \& Osterbrock (1981), and Feldman et al. (1982) is clearly 
inappropriate for the majority of the Seyfert galaxy population and should be 
abandoned.

\subsubsection{Line Asymmetry}

Of course, the FWHM is the crudest, first-order characterization of the line
profile.  Actually, the shapes of the emission lines in most emission-line
nuclei, when examined with sufficient spectral resolution (e.g., Heckman et al.
1981; Whittle 1985a; Veilleux 1991; Paper~IV), deviate far from simple 
symmetric functions (such as a Gaussian), often exhibiting weak extended wings
and asymmetry.  In fact, most Seyfert nuclei have asymmetric narrow lines, and
there seems to be a preponderance of blue wings, usually interpreted as
evidence of a substantial radial component in the velocity field coupled with
a source of dust opacity.  It would be highly instructive to see if this trend
extends to LINERs and transition objects, as it could offer insights into 
possible differences between the NLRs of objects with high and low ionization.

The majority of the objects in our survey have emission-line spectra of 
adequate signal-to-noise ratio (S/N) that possible profile asymmetries can be 
discerned (see Fig.~9 in Ho 1996 and additional examples in Paper~IV).  
Since the red spectra of our survey have higher dispersion than the blue 
spectra (FWHM resolution $\sim$100 \kms\ vs. 225 \kms; see Paper~III), we 
will work with the red spectra.  While formalisms have been developed to 
quantify line asymmetries (e.g., Whittle 1985a), here we take a simpler 
approach.  All the red spectra were visually examined and assigned an 
``asymmetry code'' according to the profile shape of the H\al, \nii, and \sii\ 
lines: ``B'' (blue), ``R'' (red), and ``S'' (symmetric).  Ambiguous cases, or 
those with low S/N, were excluded.  At the resolution of the Palomar survey, 
and for the typical velocity dispersions of our galaxies, the individual 
components of the H\al+\nii\ complex and the \sii\ doublet have well-separated 
peaks.  For objects with adequate S/N, the sense of the asymmetry is generally 
noticeable on the profile at 80\% of the peak intensity, or less.  

The majority of Seyferts (\gax 90\%) have sufficient S/N to be classified.  The 
results are as follows: 29\% S, 46\% B, and 25\% R.  These percentages remain 
essentially unchanged for the spiral subsample (31\% S, 50\% B, and 19\% R).
Quite remarkably, LINERs show virtually identical statistics.  For the objects 
that are classifiable ($\sim$75\%), we find 30\% S, 46\% B, and 24\% R (all 
Hubble types) and 26\% S, 58\% B, and 16\% R (spirals only).   Blue asymmetric 
profiles are preferentially seen in both types of objects.  Transition objects 
seem to depart from this trend.  Among the 75\% of the sample that can be 
studied, most show symmetric profiles and there is no obvious preference for 
blue or red asymmetry (all Hubble types: 52\% S, 25\% B, and 23\% R; spirals 
only: 56\% S, 26\% B, and 18\% R).  We consider the results for transition 
objects somewhat less certain because their narrower lines (\S~2.4.1) make 
it more difficult to notice profile asymmetries.  Moreover, if a significant 
portion of the line core in transition objects comes from \hii\ regions, 
asymmetries from the AGN component, if present, would be most readily 
detectable in the wings of the profile, which are much more dependent on S/N.  
(Again, there are no gross differences between the Sb subsample and the full 
spiral sample.)

To summarize: LINERs and Seyferts exhibit similar trends in their narrow-line 
asymmetries, and when present, the sense of the asymmetry is preferentially to 
the blue.
 
\subsubsection{Comparison of Profiles for Different Lines}

Detailed studies of Seyferts (e.g., De~Robertis \& Osterbrock 1984, 1986) and 
LINERs (Filippenko \& Halpern 1984; Filippenko 1985; Filippenko \& Sargent 
1988; Ho et al. 1993a, 1996; Barth et al. 2001) have found that the widths of 
the forbidden lines correlate positively with their critical densities.  This 
empirical trend has been interpreted as evidence that the NLRs of these 
objects contain a wide range of gas densities (10$^2$--10$^7$ \cc), stratified 
such that the denser material is located closer to the center.  In such a 
picture, \oi\ \lamb 6300 ($n_{\rm crit}\, \approx\,10^6$ \cc) should be 
{\it broader} than \sii\ \lamb\lamb 6716, 6731 ($n_{\rm crit}\,\approx$ 3\e{3} 
\cc).

Among the objects with securely determined FWHM  for \oi\ and \sii, 
approximately 15\%--20\% of LINERs and 10\% 

\vskip 0.3cm

\begin{figure*}[t]
\centerline{\psfig{file=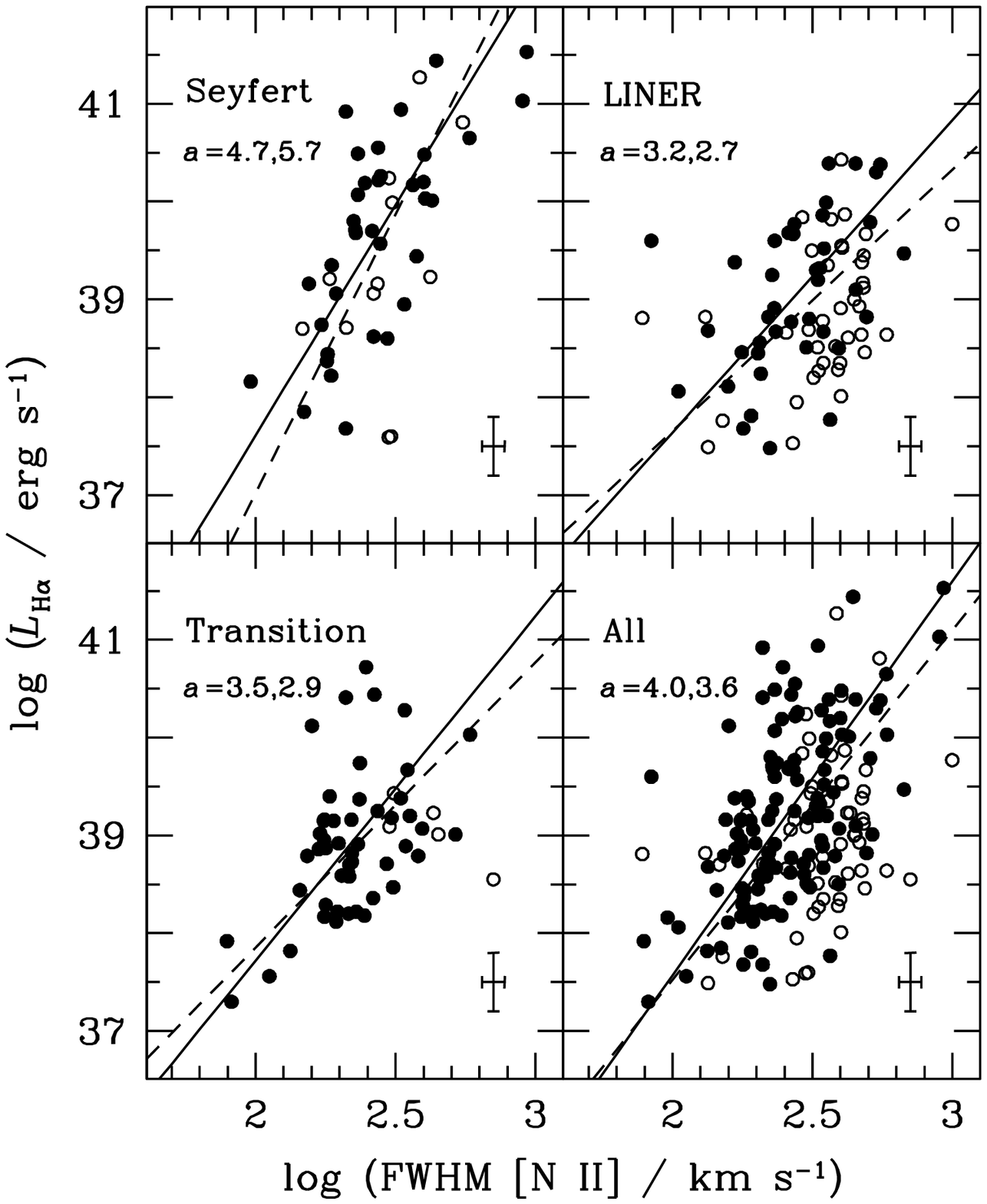,width=15.0cm,angle=0}}
\figcaption[fig8.ps]{
Correlation of FWHM of [N~II] \lamb 6583 with the luminosity of the
narrow H\al\ line.  The line widths have been corrected for instrumental
resolution, and the line luminosities have been corrected for extinction.  The
data are shown in open symbols for the entire sample and in solid symbols for
the sample of spirals.  In each panel, the linear regression line is
plotted; its slope is given in the top left corner.  The solid line
and the slope listed first correspond to the spiral subsample; the
dashed line and the slope listed second refer to the entire sample.
Representative error bars are given at the lower right corner of each panel.
\label{fig8}}
\end{figure*}
\vskip 0.3cm

\noindent
of Seyferts show evidence
of density stratification in the sense that FWHM(\oi) $>$ FWHM(\sii) (see 
Fig.~9 in Ho 1996 for an example).  In no instance is \oi\ ever observed to be
narrower than \sii.  However, these numbers need to be regarded with 
caution.  They do {\it not}\ imply that objects failing to show such profile 
differences lack density stratification, since a number of effects can 
conspire to hide this observational signature (Whittle 1985c).  Furthermore, 
our ability to discern such subtle profile differences depends strongly on 
the S/N and resolution of the data, and undoubtedly many objects have escaped 
notice because of this observational selection effect.
 
Whittle (1985c) found that Seyfert 1 nuclei have a greater likelihood of
showing profile differences in their forbidden lines than Seyfert 2 nuclei.  
The implication is that somehow density stratification in the NLR is directly
related to the presence of a broad-line region.  In the present sample, 
the same trend seems to hold (see also Ho et al. 1993a), in that, among those 
objects with detectable profile differences between \oi\ and \sii, $\sim$50\% 
of the LINERs and $\sim$80\% of the Seyferts have broad H\al\ emission, 
significantly higher than the respective detection rates of broad H\al\ in the 
whole sample (Paper~IV).  But, once again, this result is difficult to 
interpret, since selection effects heavily favor the detection of both of 
these traits in objects having data of high S/N. 

\begin{figure*}[t]
\centerline{\psfig{file=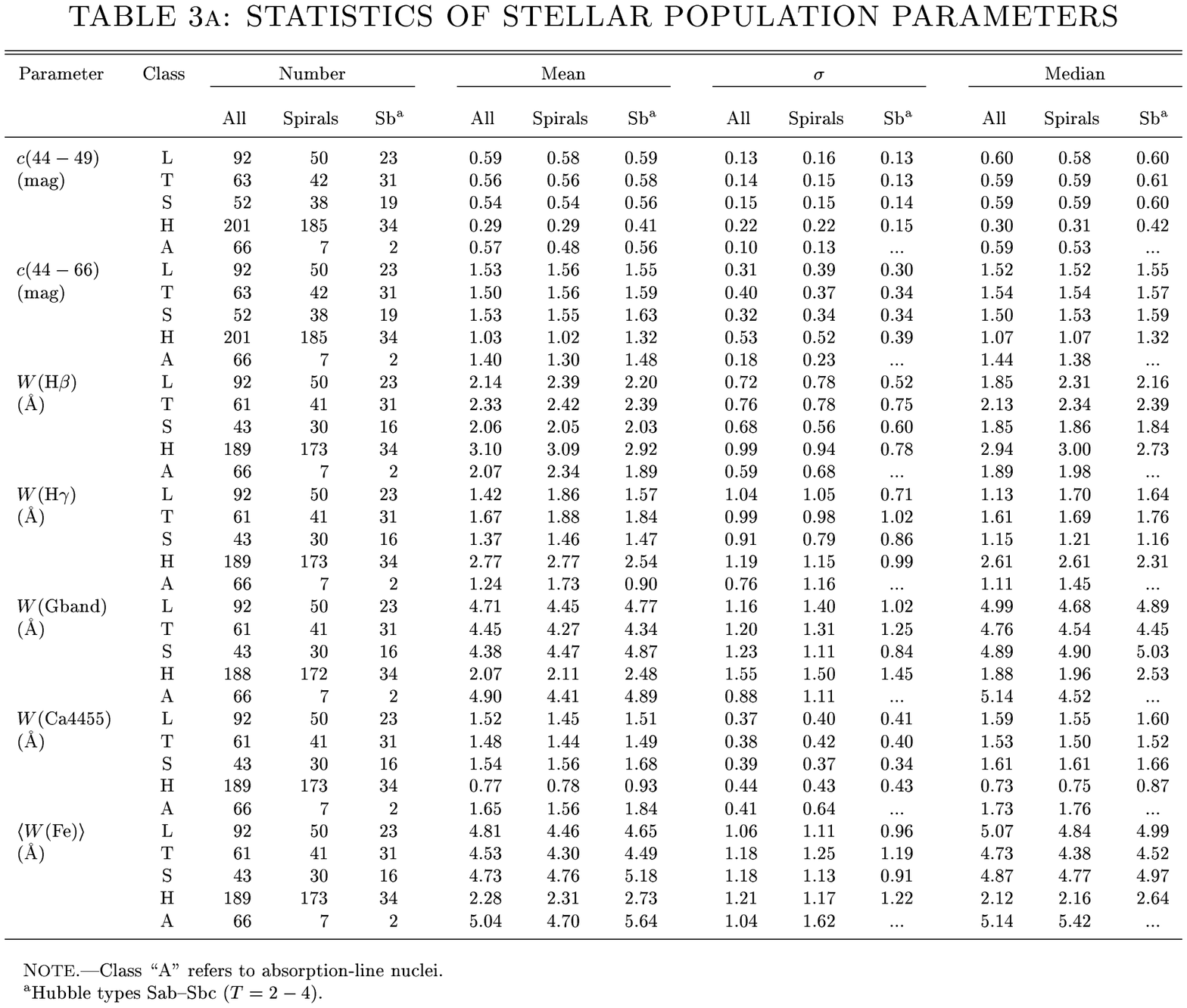,width=18.0cm,angle=0}}
\end{figure*}

\subsection{Nuclear Stellar Content}

Paper~III gives several parameters that are useful indicators of the nuclear 
stellar population, at least in a statistical and relative sense.  These are 
summarized in Table~3{\it a}, and comparisons between the different classes of 
objects are given in Table~3{\it b}.  To a first approximation, the three AGN 
subtypes reveal no statistical differences in their nuclear stellar content, 
especially after isolating the spiral and Sb subsample.  This is reflected in 
the spectrophotometric colors [$c(44-49)$ and $c(44-66$] and in the Balmer 
and metal-line absorption features.  Figure~9 plots the age-sensitive indices, 
$W$(H\bet) and $W$(H$\gamma$) against the metallicity-sensitive indices, 
$\langle W$(Fe)$\rangle$ and $W$(Ca4455).  All three classes of AGN
nuclei overlap considerably, and there are no noteworthy differences 
among them
%
%
(Table~3{\it b}).  We have also examined the behavior of the G band
near 4300 \AA, and it, too, is very homogeneous among the three classes.
 
As expected, \hii\ nuclei occupy a distinctly different locus with respect to
the AGN nuclei.  They have strong Balmer lines and weak metal lines --- 
signatures of a young to intermediate-age stellar population --- consistent 
with the expectation that the emission-line spectrum of \hii\ nuclei is 
powered predominantly by young, massive stars.

Table~3{\it a}\ also lists the statistics for the absorption-line nuclei 
(nuclei with no detectable emission lines) from the Palomar survey.  Most of 
the galaxies are ellipticals and lenticulars, which have characteristically old 
stellar populations.  The statistics for the spiral and Sb subsamples are 
limited, but they do not appear to be grossly dissimilar from those of the 
whole sample.  It is striking that both the absorption-line nuclei and 
the three classes of active nuclei have such closely matching stellar 
population parameters.  We can infer, from this comparison, that an old 
stellar population prevails in most LINER, transition, and Seyfert nuclei.

\begin{figure*}[t]
\centerline{\psfig{file=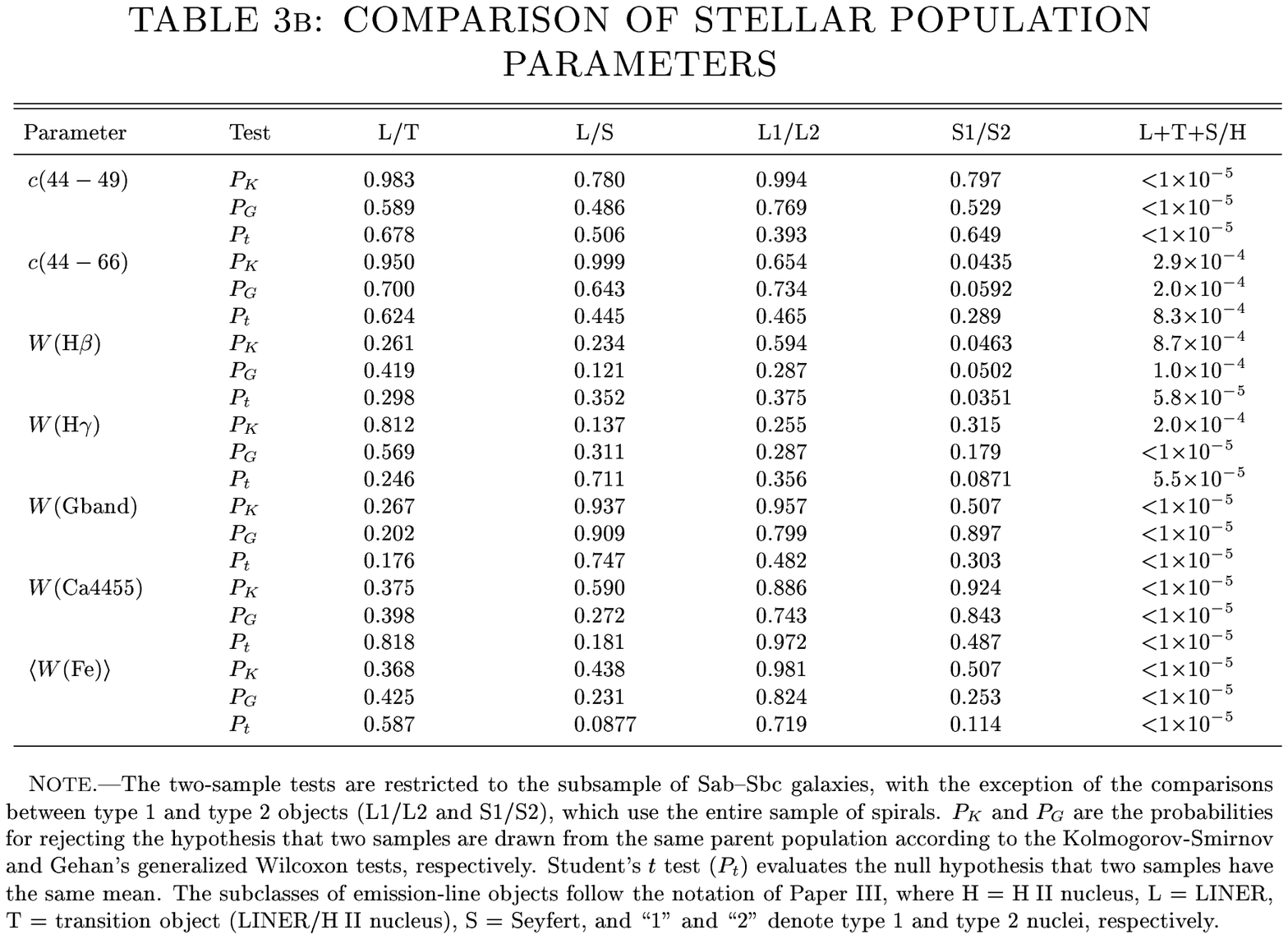,width=18.5cm,angle=0}}
\end{figure*}

\section{Discussion}

\subsection{The Origin of the Low-Ionization State in LINERs}

While there is now abundant evidence that a significant fraction of LINERs are 
accretion-powered sources [see recent reviews by Ho (1999a, 2002) and Barth 
(2002)], it is still unclear what physical parameters actually determine the 
low-ionization state in LINERs.  In the context of AGN photoionization, the 
optical signature of LINERs has been primarily attributed to a low ionization 
parameter, $U$, defined as the ratio of the density of ionizing photons to the 
density of nucleons at the illuminated face of a cloud.  Whereas the NLR 
spectrum of Seyferts can be well reproduced with $\log U \approx -2.5\pm0.5$ 
(e.g., Ferland \& Netzer 1983; Stasi\'nska 1984; Ho, Shields, \& Filippenko 
1993b), that of LINERs requires $\log U \approx -3.5\pm1.0$ (Ferland \& Netzer 
1983; Halpern \& Steiner 1983; P\'equignot 1984; Binette 1985; Ho et al. 
1993a).  What factors contribute to the lower ionization parameters in LINERs?

The ionization parameter is conventionally related to the physical parameters 
of a line-emitting region by the expression $U = Q_{\rm H}/(4 \pi r^2 n c)$, 
where $Q_{\rm H}$ is the number of ionizing photons s$^{-1}$, $n$ is the gas 
number density, and $r$ is the distance between the central ionizing source 
and the illuminated cloud.  Recasting the structure of the nebula in terms of 
a volume filling factor $\epsilon$, $U \propto (Q_{\rm H} n \epsilon^2)^{1/3}$. 
Unfortunately, our observations do not constrain $\epsilon$.   We can, however, 
estimate the remaining two variables, since $Q_{\rm H} \propto 
L_{{\rm H}\alpha}$ in an ionization-bounded nebula and $n \approx n_e$.   From 
\S~2.3, we know that LINERs are intrinsically less powerful sources than 
Seyferts; the typical H\al\ luminosities differ by approximately an order of 
magnitude.  On the other hand, the electron densities in LINERs are lower than 
those in Seyferts by a factor of $\sim$3.   Thus, neglecting for the moment 
possible systematic variations in $\epsilon$ between the two classes, we 
expect LINERs to have values of $U$ that are, on average, lower by 
$\sim 30^{1/3}$, or $\sim$3.  While this does not fully reconcile the factor
of 10 difference in $U$ between LINERs and Seyferts, it is the first direct
demonstration that systematic variations in nebular conditions may be
responsible for the spectral distinction between low-ionization and
high-ionization AGNs.

Clearly, further progress would require data on $\epsilon$, especially in view
of its relatively strong influence on $U$.  For the central photoionization
picture to remain viable, the emission nebulae in LINERs should have lower
volume filling factors than in Seyferts, on average by a factor of a few.
Observational constraints on $\epsilon$ can be derived from knowledge of the
size of the NLR, in conjunction with its line luminosity and density.  The 
ground-based narrow-band imaging surveys of Keel (1983b) and Pogge (1989a), 
which included a number of nearby LINERs, find that the majority of them
have fairly compact, centrally concentrated NLRs.  The emission cores are 
generally unresolved or only marginally resolved under 1\asec--2\asec\ seeing.
This result has been verified for a limited sample of 14 LINERs imaged with 
the {\it Hubble Space Telescope (HST)}: the bulk of the line emission is 
confined to size scales of tens to hundreds of parsecs (Pogge et al. 2000).  
The compact morphology of the NLRs in LINERs ostensibly seems to differ from 
the extended emission-line structures and ionization cones commonly associated 
with Seyfert galaxies (e.g., Pogge 1989b; Mulchaey, Wilson, \& Tsvetanov
1996).  However, one must regard this comparison with some caution.  Past
emission-line

\vskip 0.3cm

\begin{figure*}[t]
\centerline{\psfig{file=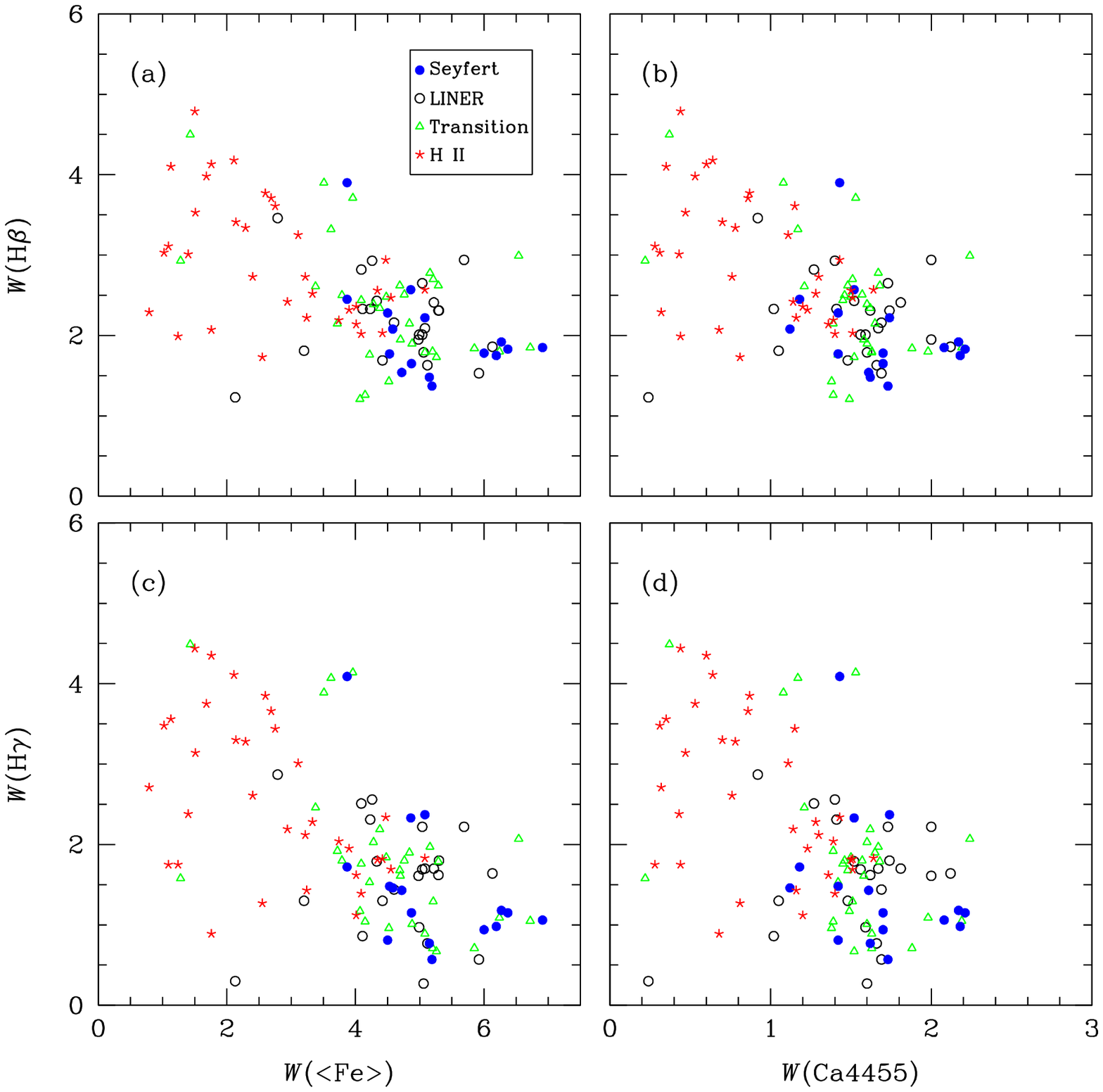,width=15.0cm,angle=0}}
\figcaption[fig9.ps]{
Stellar absorption indices indicative of the age [$W$(H\bet) and
$W$(H$\gamma$)] and metallicity [$\langle W$(Fe)$\rangle$ and $W$(Ca4455)] of
the stellar population, as defined in Paper III, for Seyferts (solid circles),
LINERs (open circles), transition objects (triangles), and \hii\ nuclei
(stars).  Only Sab--Sbc galaxies are included.
\label{fig9}}
\end{figure*}
\vskip 0.3cm

\noindent
imaging studies have targeted Seyferts that are generally much 
more luminous than the ``garden-variety'' sources represented in the Palomar 
survey.  If the linear extent of the NLR correlates with line luminosity, as 
found by Mulchaey et al.  (1996), it is possible that the majority of nearby 
Seyferts have more compact NLRs than previously believed.  The work of Keel 
(1983b) tentatively supports this hypothesis.  The majority of the galaxies in 
Keel's study overlap with the Palomar survey, and thus can be 
spectroscopically classified according to our system.  In all, there are six 
Seyferts and 15 LINERs, and the average seeing-corrected FWHM sizes of the 
H\al+\nii\ emission regions are, respectively, 4.4\asec\ and 3.5\asec.  These 
statistics are clearly too limited to be definitive, but they illustrate that 
the NLRs of nearby (low-luminosity) Seyferts may indeed be quite compact.  It 
would be desirable to address these issues by conducting a high-resolution 
imaging survey of the Palomar LINERs and Seyferts matched in line luminosity 
and Hubble type.

A common theme echoed throughout this paper is that LINERs and Seyferts share, 
perhaps surprisingly, a large number of traits.  After factoring out slight 
differences in Hubble type distribution, the global properties of their 
host galaxies are virtually identical.  The distinguishing features that 
emerge all pertain to the circumnuclear environment: LINERs have lower gas 
densities and less internal reddening relative to Seyferts.  This suggests 
that the central regions of LINERs have characteristically lower amounts of 
cold interstellar material.  It is reasonable to speculate that the lower 
circumnuclear gas content may lead to a reduction in the gas reservoir 
available to fuel the central engine, which in turn explains the observed 
lower luminosity output.

A sizable reduction in accretion rate can have other consequences.  Several 
recent studies have remarked that the broad-band spectral energy distributions 
of low-luminosity AGNs, particularly LINERs, are systematically different from 
those of higher luminosity sources such as classical Seyfert 1 nuclei and 
quasars (Ho 1999b, 2002; Ho et al. 2000).  The most salient feature is a 
marked deficit of optical and ultraviolet (UV) photons normally attributed to 
thermal emission from an optically thick, geometrically thin accretion disk. As 
suggested by Ho (2002), this modification of the spectral energy distribution 
hardens the ionizing radiation field, which, all else being equal, boosts the 
strengths of the low-ionization lines.  This effect, however, is likely to 
be only secondary compared to the variation in ionization parameter.

\subsection{The Role of Shocks}

The relevance of shocks to the excitation of LINERs has been a topic of 
ongoing debate ever since LINERs were first identified (Koski \& Osterbrock 
1976; Fosbury et al. 1978; Heckman 1980; Dopita \& Sutherland 1995; 
Alonso-Herrero et al. 2000; Sugai \& Malkan 2000).  Optical emission-line 
diagnostics, unfortunately, do not discriminate well between shocks and 
conventional AGN photoionization.  Although the UV region offers more promise, 
according to the fast-shock ($\upsilon \approx 150-500$ \kms) models of Dopita 
\& Sutherland (1995), to date none of the LINERs that have been studied 
spectroscopically in the UV using the {\it HST}\ have revealed the predicted 
strong high-excitation UV lines (see summary in Ho 1999a, and references 
therein)\footnote{The case discussed by Dopita et al. (1997) concerns the 
circumnuclear {\it disk}\ of M87, not the nucleus itself.  Sabra et al. (2002) 
show that the UV--optical spectrum of the nucleus of M87 is best explained by 
a multi-component photoionization model.}.
The present work has some relevance to the issue of shocks.  From analysis 
of the line profiles (\S~2.4), we find that the nuclear ionized gas in LINERs 
and Seyferts have comparable velocity dispersions; depending on Hubble type, 
100 \lax\ $\upsilon_{\rm gas}$ \lax\ 170 \kms, where $\upsilon_{\rm gas}$ = 
0.5 FWHM(\nii).  Now, in spiral galaxies at least part of the line width 
observed through a ground-based aperture must come from spatially unresolved 
rotation of the inner disk, and so the true random, cloud-cloud impact 
velocities should be less than $\upsilon_{\rm gas}$.  Even so, these velocities
already fall considerably short of the values assumed in fast-shock models.  
Moreover, if LINERs are preferentially shock excited, one would naively expect 
their internal gas motions to be higher than in Seyferts, contrary to what is 
observed.  In fact, judging by the frequency with which asymmetric narrow-line 
profiles are observed in LINERs, as well as the clear preference
for the asymmetry to occur blueward of the line center, the bulk velocity 
field of their NLRs appears to be remarkably similar to that of Seyferts.

\subsection{The Role of Stellar Photoionization and the Nature of 
Transition Objects}

Beginning with the work of Terlevich \& Melnick (1985), there have been a 
number of attempts to invoke photoionization by hot, young stars as the 
primary source of excitation for LINERs and related objects.  As with the 
shock models discussed above, the main motivation for the stellar-based models 
is clear: if alternatives to AGN photoionization can be found to give a 
satisfactory explanation of LINERs, then LINERs should not be regarded as AGNs.
Filippenko \& Terlevich (1992) and Shields (1992) showed that the primary 
optical spectral features of transition objects\footnote{Filippenko \& 
Terlevich (1992) and Shields (1992) refer to these sources as ``weak-[O~I]  
LINERs,'' but they are equivalent to transition objects.} can be reproduced by 
photoionization by O-type stars having effective temperatures \gax 45,000 K 
embedded in an environment with high density and low ionization parameter. 
Taniguchi, Shioya, \& Murayama (2000) advocate that the ionization source for 
LINERs can be supplied by a cluster of planetary nebula nuclei formed 100--500 
Myrs following a nuclear starburst. 

An improved treatment of the problem studied by Filippenko \& Terlevich (1992) 
and Shields (1992) was recently presented by Barth \& Shields (2000), who 
modeled the ionizing source not as single O-type stars but as a more realistic 
evolving young star cluster.  Barth \& Shields confirm that young, massive 
stars can indeed generate optical emission-line spectra that match those of 
transition objects, but only under the following conditions: the cluster needs 
to be formed in an instantaneous burst,  its metallicity should be solar or 
greater, and its age must lie in the narrow range $\sim$3--5 Myr.  The latter 
restriction ensures that there are sufficient Wolf-Rayet (W-R) stars available 
to supply the extreme-UV photons necessary to boost the low-ionization lines.  
As Barth \& Shields emphasize, however, simple considerations of the 
demographics of emission-line nuclei observed in the Palomar survey (Paper~V) 
indicate that the starburst model is unlikely to apply to most transition 
objects, especially those hosted by earlier-type galaxies.  

The models of Barth \& Shields (2000) also succeed in explaining the optical 
spectra of bona fide LINERs, again provided that the starburst is caught 
during the brief phase when W-R stars exist.  To achieve consistency with the 
relative strengths of \oi, \nii, and \sii\ observed, the models additionally 
require above-solar metallicities and the coexistence of a high-density and a 
low-density component, similar to the scenario envisioned by Shields (1992).  
As with the transition objects, however, the starburst model for LINERs 
appears to be applicable only to late-type galaxies, which host only a 
small fraction of the known LINERs.

As recognized by Barth \& Shields (2000), the critical role played by W-R 
stars in their model presents a somewhat perplexing problem: why 
do LINERs and transition objects almost never show W-R features (e.g., the 
broad ``bump'' near 4650 \AA) in their spectra?  Conversely, why do galaxies 
with detected W-R features (``W-R galaxies''; see Conti 1991) seldom qualify 
as LINERs or transition objects according to their narrow-line spectra?   
Since W-R features are recognized most commonly, and perhaps selectively, 
in late-type galaxies, Barth \& Shields (2000) speculate that the absence of 
LINER-like spectra in W-R galaxies may be a consequence of their low
metallicity, and possibly low density.  They attribute the apparent rarity of 
the W-R bump in LINERs and transition nuclei to the difficulty of detecting 
this weak, broad feature in the presence of strong starlight contamination 
from the bulge of the host galaxy.  We do not believe this to be the case.
While contamination from the underlying bulge stars indeed does pose a serious 
challenge to detecting weak emission lines in galactic nuclei, all 
emission-line measurements from the Palomar survey were done {\it after}\ 
careful starlight subtraction (Paper III).  To be sure, the correction for 
starlight was not perfect in all cases, especially for objects with extremely 
weak emission lines.  Nonetheless, H\bet\ emission is detected unambiguously in 
the vast majority of the LINERs and transition nuclei in the Palomar survey.
Furthermore, the starlight-subtracted spectra in the region of the W-R bump 
($\sim 4600-4700$ \AA) have residuals comparable to that of the continuum near 
H\bet.  For high (\gax\ solar) metallicities, the strength of the W-R bump is 
expected to be comparable to, if not greater than, that of H\bet\ (Schaerer \& 
Vacca 1998).  It is thus hard to imagine how it could have been missed, 
especially considering the size of the Palomar sample ($\sim$160 LINERs and 
transition objects).

Are there other indications that low-ionization nuclei contain young stars?  
The two best examples are NGC 404 (a LINER) and NGC 4569 (a transition object).
Both objects have a prominent nuclear star cluster detected in {\it HST}\ UV 
images (Maoz et al. 1995; Barth et al. 1998), and follow-up spectroscopy 
reveals that the UV light is dominated by emission from massive, young stars 
(Maoz et al. 1998).  Although the youth of the nuclear stellar population in 
NGC 404 and NGC 4569 manifests itself most dramatically in the UV, it is also 
unmistakable in the optical.  The blue Palomar spectra of these two objects 
(Paper II) show extremely prominent H$\gamma$ and H\bet\ absorption lines. 
We measure $W$(H$\gamma$) = 3.4 and 4.5 \AA\ and $W$(H\bet) = 3.6 and 4.5 \AA\ 
for NGC 404 and NGC 4569, respectively.  The metal lines are correspondingly 
weakened by dilution from the continuum of hot stars.  The respective indices 
for NGC 404 and NGC 4569 are $W$(Gband) = 1.8 and 0.74 \AA, $W$(Ca4455) = 0.68 
and 0.37 \AA, and $W$($\langle {\rm Fe} \rangle$) = 2.3 and 1.4 \AA.  
The vast majority of LINERs and transition nuclei, however, do {\it not}\ 
resemble NGC 404 and NGC 4569.  UV-bright nuclei, whether stellar or 
nonstellar, occur in only $\sim$20\%--25\% of nearby AGNs (Maoz et al. 1995; 
Barth et al. 1998).  More striking still, as shown in \S~2.5, active nuclei 
in nearby galaxies, irrespective of spectral classification, contain 
predominantly {\it old}\ stellar populations.  The optical stellar indices 
of NGC 404 and NGC 4569 (given above) are highly unrepresentative of the bulk 
of LINERs and transition nuclei of similar Hubble type (see Table~3{\it a}).  
 
How confident are we that young stars cannot be present in significant numbers?
Perhaps the number of young stars required to drive the emission lines would
be imperceptible in the presence of the more abundant old stars.  To evaluate 
this possibility, consider the following.  The average H\al\ luminosity of 
LINERs and transition objects, $\sim$3\e{39} \lum, if generated under Case B 
recombination, corresponds to an ionizing photon rate of $\sim$3\e{51} 
s$^{-1}$.  If the ionization can be entirely attributed to an ongoing  
starburst, say at an age of 5 Myr, we estimate from the models of Leitherer 
et al. (1999) that the associated stellar component has an absolute 
$B$-band magnitude of $M_B \approx -14.5$ mag.  We have chosen the calculations
that assume solar metallicity and a Salpeter stellar initial mass function 
with an upper-mass cutoff of 100 \solmass.  This can be directly compared to 
the nuclear optical magnitudes compiled in Paper III, which were derived from 
our spectrophotometric measurements.   The average monochromatic absolute 
magnitude at 4400 \AA\ is $\langle M_{44} \rangle = -16.1$ mag for LINERs and
$\langle M_{44} \rangle = -15.5$ mag for transition nuclei.  The starburst 
component, if present, would comprise $\sim$25\%--40\% of the observed 
$B$-band light, and thus should be readily detectable.

The above considerations cast doubt on the general applicability of models 
that seek to account for the excitation of LINERs and transition nuclei using 
young stars.  Models that rely on a starburst rich in W-R stars, such as that 
by Barth \& Shields (2000), face the puzzle that W-R stars are rarely detected 
in these systems.  Post-starburst models (e.g., Taniguchi et al. 2000) bypass 
this problem, but like the starburst models, they cannot escape the predicament
that the nuclear stellar population is demonstrably old.  

Ho et al. (1993a) originally proposed that transition objects are composite in 
nature, namely systems consisting of a ``normal'' LINER nucleus whose signal 
has been diluted by neighboring, circumnuclear \hii\ regions, or simply by 
\hii\ regions projected along the line of sight into the spectroscopic 
aperture.  This seems to be the most natural explanation for their location, 
sandwiched between \hii\ regions and LINERs, in optical line-ratio diagrams. 
A similar argument, based on decomposition of line profiles, has been made by 
V\'eron, Gon\c{c}alves, \& V\'eron-Cetty (1997) and Gon\c{c}alves, 
V\'eron-Cetty, \& V\'eron (1999).   As noted in Paper~V and in \S~2.1, 
the overall distribution of Hubble types for transition objects tends to be 
skewed toward later morphologies.  This is not unexpected, to the extent that 
star formation is more prevalent in later type galaxies.  More intriguingly, 
we find that the host galaxies of transition nuclei seem to exhibit 
systematically higher levels of recent star formation compared to LINERs of 
{\it matched}\ morphological types (i.e., within the Sb subsample).  This is 
suggested by the higher $L_{\rm FIR}/L_B^0$ ratios and mildly bluer broad-band 
optical colors in transition objects.  Since these are spatially unresolved 
measurements, however, we do not know the location of the enhanced star 
formation.  Moreover, we showed that the host galaxies of transition nuclei 
also have a tendency to be slightly more inclined than LINERs.  Thus, all 
else being equal, transition-type spectra seem to be found precisely in those 
galaxies whose nuclei have a high probability of being contaminated by 
extra-nuclear emission from star-forming regions. 

The above proposition, it may be argued, seems to be at odds with the stellar 
population of transition nuclei.  If star-forming regions contribute 
significantly to the integrated nuclear emission of transition objects, then 
why is there no strong evidence that they have a younger population than 
LINERs\footnote{The stellar population parameters of transition nuclei indeed 
{\it do}\ indicate systematically younger ages than LINERs (Table 3{\it a}), but the 
differences are not large enough to be statistically significant.}?  
This apparent inconsistency perhaps can be resolved by recognizing that 
in giant \hii\ regions the line-emitting regions can be considerably more 
extended than, and often offset from, the stellar continuum sources (see, e.g., 
Whitmore et al. 1999; Maoz et al. 2001).  A spectroscopic aperture, therefore, 
can intercept significant line emission  from near-nuclear or projected \hii\ 
regions without admitting much of the associated continuum light.

\subsection{The Role of Interactions}

The influence of the local environment on nuclear activity is not the primary 
subject of this investigation, but our analysis touches upon a few items 
relevant to this issue.  Although tidal interactions have been well 
demonstrated to have an important influence on nuclear star formation, their 
influence on AGN activity, albeit often implicated in the literature, is far 
from clear.  This is especially true of lower luminosity AGNs such as 
Seyfert nuclei (see Combes 2001, and references therein).  

Statistical studies of this type are at the mercy of selection effects and 
sample biases.  In this respect, the Palomar survey has a number of merits, 
as emphasized by Ho \& Ulvestad (2001). Schmitt (2001), taking advantage of 
this fact, has studied the frequency of companions for galaxies with different 
nuclear types in the Palomar survey.  The approach taken by Schmitt is very 
similar to ours.  He evaluates the influence of companion galaxies using 
$\rho_{\rm gal}$ taken from Paper III and a parameter equivalent to, but 
quantitatively slightly different from, $\theta_p$.  A galaxy is considered to 
have a companion if another galaxy with a magnitude difference of $\pm$3 mag 
and a velocity difference of $\pm$1000 \kms\ is found within a separation of 
5 $D_{25}$ of the primary.  By either measure, Schmitt concludes that 
the local environment has no correlation with the activity type, as long as one
properly accounts for the morphology-density relation.  This is precisely what 
we found in \S~2.2.  Thus, although non-axisymmetric perturbations from 
galaxy interactions may be effective in driving gas from galactic scales 
($\sim$10 kpc) to the circumnuclear region ($\sim 0.1-1$ kpc), wherein nuclear
starbursts may ignite (e.g., Barnes \& Hernquist 1991), the gas evidently has 
more trouble dissipating to scales pertinent to feeding a central black hole 
(\lax 1 pc).  Ho et al. (1997d) arrived at the same conclusion in considering 
the effect of bars on nuclear activity found in the Palomar survey.  Schmitt 
notes that the null effect of companions on AGN fueling cannot be easily 
dismissed by appealing to the low level of activity represented by the Palomar 
objects.  The frequency of companions for the Palomar Seyferts appears to be 
comparable to that in the much more luminous sample studied by Schmitt et al. 
(2001).

\subsection{Seyfert Galaxies: Not So Unified?}

According to the simplest version of AGN unification (see, e.g., Antonucci 
1993; Wills 1999), type 2 Seyferts are intrinsically the same as type 1 
Seyferts, but viewed from an angle in which a small-scale dusty ``torus'' 
obscures its broad-line region.  For this picture to hold, we expect both 
types to have similar isotropic properties.  As before, here it is of 
paramount importance to minimize potential selection effects, which 
traditionally have plagued analyses of this kind. The Palomar sample is 
again quite valuable in this regard.  Its main disadvantage is the small 
size; when the analysis is confined to the spiral galaxies, there are only 
18 Seyfert 1s and 20 Seyfert 2s.  Nonetheless, some trends are apparent.

As can be seen from Table 1{\it b}, the host galaxy parameters of both types of 
Seyferts are mostly well matched, with the following exception.  Relative to 
Seyfert 2s, Seyfert 1s have comparable FIR emission (normalized to the 
optical) but {\it hotter}\ FIR colors (higher $S_{60}/S_{100}$ and 
$S_{25}/S_{60}$ ratios).   Models of optically thick obscuring tori 
(e.g., Pier \& Krolik 1992; Efstathiou \& Rowan-Robinson 1995) predict
a significant degree of anisotropy for the infrared emission, such that 
face-on tori should be hotter than edge-on tori.  The systematic differences 
in FIR colors are therefore in qualitative agreement with the model 
predictions.  Heckman (1995), by comparing the ratio of 10 $\mu$m emission to 
1.4~GHz and \oiii\ \lamb 5007 luminosity in broad-line versus narrow-line 
AGNs, also concluded that the mid-infrared emission must be emitted 
anisotropically.

In terms of nuclear properties, we find that Seyfert 1s tend to have weaker 
stellar indices than Seyfert 2s.  This result is simple to understand in 
terms of dilution by the stronger featureless continuum in Seyfert 1s, because 
relative to Seyfert 2s they have a more dominant, directly viewed AGN 
component compared to the underlying galaxy.  Although both Seyfert types have 
similar narrow-line H\al\ luminosities (Table 2{\it b}), the total (narrow 
plus broad) line luminosity of Seyfert 1s is higher than that of Seyfert 2s.  
Thus, this result does not conflict with the unified model either.

Two points, however, are less straightforward to grasp.  Taken at face value, 
they appear to violate the simplest formulation of the unified model.   First, 
as mentioned in \S~2.2, Seyfert 1s seem to prefer environments of higher 
galaxy density than Seyfert 2s.  Since both subsamples are well matched in 
Hubble type and total galaxy luminosity, we cannot dismiss it by appealing to 
the morphology-density effect.  Second, Seyfert 1s have statistically higher 
NLR electron densities than Seyfert 2s.  Their density distributions differ at 
the level of $P_K$ = 3\% and $P_G$ = 1\%, while the difference in their means 
(632 and 380 \cc, respectively) is significant at the level of $P_t$ = 0.8\%.

The following complication, however, might obviate the above apparent 
inconsistency with the unified model.  The spectral classifications in the 
the Palomar survey, as in all large spectroscopic surveys, are based on 
integrated-light spectra, with no means of distinguishing scattered emission 
from direct emission. The Palomar survey contains a large number of objects 
with very weak broad lines, many classified as type 1.8 or type 1.9 sources.  
For these objects, we do not know for sure whether the broad-line emission is 
viewed directly or is largely reflected into our line of sight.  If the latter 
is true, as in the case of NGC 1068 (Antonucci \& Miller 1985), then such 
objects should be classified as type 2 rather than type 1 objects.  Sensitive 
spectropolarimetric observations of the Palomar Seyferts are needed before 
they can be used to definitively test the unified model.

\subsection{What ``Activates'' Galactic Nuclei?}

Dynamical studies suggest that most, perhaps all, galactic bulges contain 
massive black holes (Magorrian et al. 1998; Gebhardt et al. 2002).  Insofar 
as AGNs signify black hole accretion, AGN surveys can serve as an alternative 
and efficient tool to assess black hole demographics.   As noted by Ho (2002), 
the high AGN fraction among bulge-dominated galaxies in the Palomar survey 
(\gax 50\%--75\% for E--Sbc galaxies) supports the idea that 
black holes are ubiquitous in bulges.  What about the small, but still 
sizable, minority of bulged galaxies that are classified as \hii\ nuclei?   
Why are they ``inactive''?  One possibility is that they lack black holes.  A 
more mundane alternative, however, is that the AGN signal is simply swamped by 
the much more dominant light from the \hii\ regions.  Among Sb galaxies, for 
instance, the H\al\ luminosity of \hii\ nuclei easily rivals that of Seyferts 
and typically exceeds that of LINERs and transition objects by an order of 
magnitude.  If this scenario is correct, we expect that spectra taken at 
high spatial resolution would look progressively more AGN-like.

Table 1 shows that the host galaxies of \hii\ nuclei tend to be slightly 
richer in atomic hydrogen compared to the host galaxies of AGNs, although we 
know little of the internal distribution of the gas. The higher gas content 
plausibly leads to elevated star formation, both on nuclear scales, as 
reflected in the spectral classification, and on galaxy-wide scales, as 
indicated by the integrated optical colors and FIR properties.  An additional 
trend is noteworthy.  For a given Hubble type, \hii\ nuclei are preferentially 
found in galaxies of somewhat lower total optical luminosity; they tend to be 
less luminous than AGN hosts by $\sim 0.3-0.4$ mag.  This is probably just a 
consequence of the inverse correlation between gas content and optical 
luminosity in spiral galaxies (Roberts \& Haynes 1994).  The systematically 
lower luminosities account for the more compact isophotal diameters (Holmberg 
1975) and reduced bulge luminosities, which depend directly on the total 
luminosities.

With the knowledge that all bulges are likely to contain massive black holes, 
and therefore the necessary condition to host AGNs, the longstanding, 
unanswered question of what triggers the activity becomes even more acute.  
Apparently, the mere availability of gas on circumnuclear scales --- inferred 
from the strong line emission and high internal reddening in \hii\ nuclei --- 
is not sufficient; most of the gas never gets accreted, hence leading to the 
low level of activity that we speculate lies masked by the \hii\ regions.  The 
key issue, then, is what factors predispose a galactic nucleus to convert most 
of its gas into stars instead of funneling it to feed the central engine.  
Environment seems to matter little, as discussed in \S~3.4.  On nuclear 
scales, the density of the gas, at least in the warm ionized phase, is not 
grossly different between AGN and non-AGN hosts.   The most disparate trait 
between the two samples is the width of the emission lines.  
Among Sb galaxies, $\langle$FWHM(\nii)$\rangle$ = 164 \kms\ for \hii\ nuclei, 
whereas $\langle$FWHM(\nii)$\rangle$ = 258 \kms\ for LINERs, transition 
objects, and Seyferts combined; the two samples differ at a very high level of 
significance ($P_K \approx P_G \approx P_t$ \lax\ $10^{-5}$; Table 2{\it b}).  
The slight difference in total or bulge luminosity between the two groups 
cannot account for such a large difference in line widths.  The gas in the 
central regions of galaxies hosting \hii\ nuclei is {\it kinematically 
colder}\ than in galaxies that host active nuclei.  Expressed in another 
way, the gas surrounding \hii\ nuclei has higher angular momentum than the 
gas in AGN nuclei.  Thus, the angular momentum content of the 
circumnuclear gas may be the critical factor that determines whether 
material can be channeled to the center for AGN fueling. 

\section{Summary}

We have used the database assembled in Paper~III to quantify statistically 
the global and nuclear properties of the various subclasses of emission-line 
nuclei found in nearby galaxies.  To mitigate spurious results that can 
arise from differences in Hubble type distribution, we have focused our 
attention on a restricted set of Sab--Sbc galaxies.  The main results 
can be summarized as follows.

\begin{enumerate}

\item{The host galaxies of LINERs, transition nuclei, and Seyferts have fairly 
uniform large-scale properties.  The most notable exception is that transition 
objects, relative to LINERs, tend to be somewhat more highly inclined and 
show mild evidence for enhanced star formation.}

\item{The nebular parameters of LINERs are broadly similar to those of 
transition nuclei, but they differ quite dramatically from those of 
Seyferts.  Seyfert nuclei tend to have significantly stronger line emission, 
denser gas, and higher levels of internal reddening. These trends suggest 
that Seyferts have more gas-rich circumnuclear environments, and plausibly 
larger accretion rates, than LINERs.}

\item{The characteristically lower nuclear luminosities and densities 
of LINERs compared to Seyferts can partly account for the difference 
in ionization parameter between these two classes of objects.}

\item{The line-emitting regions in LINERs and Seyferts share very similar 
kinematics.  This is reflected both in their line widths and profile 
asymmetries.  Transition objects, by contrast, exhibit markedly narrower 
emission lines, presumably because a greater fraction of its ionized 
gas has disklike  kinematics.  All three classes obey a correlation 
between line luminosity and line width, approximately of the form 
$L \propto {\rm FWHM}^a$, with $a \approx 3-4$.}

\item{Based on the modest gas velocity dispersions observed in LINERs and 
Seyferts, as well as their similarity between the two classes of objects, we 
argue that fast shocks are unlikely to be an important contributor to the 
excitation of LINERs.}

\item{The central regions of most nearby AGNs have uniformly old stellar 
populations.  We show that this poses a serious obstacle to models that 
invoke young stars as the primary energy source to power the emission lines.}

\item{The hypothesis that transition objects are composite LINER/\hii\ nuclei 
remains viable as long as the \hii\ region component does not contribute 
appreciably to the measured stellar continuum.}

\item{Consistent with other recent studies, we find that the local environment, 
as measured through either the local galaxy density or the distance to the 
nearest sizable companion, has a negligible effect on the general spectral 
class of emission-line nuclei (Seyferts, LINERs, or transition objects).}

\item{Type 1 and type 2 Seyferts have largely, but not completely, similar 
global and nuclear properties.  Two of the differences between the two types, 
as reflected in the local environment of the host galaxies and in the 
electron densities of the emission-line regions, may present a challenge for 
unification models for Seyfert galaxies.}

\item{The primary trait that distinguishes AGN and non-AGN host galaxies on 
small scales appears to be the velocity field of the nuclear gas. The 
line-emitting gas in \hii\ nuclei is systematically kinematically colder, 
and hence has higher angular momentum, than the gas in active nuclei.  We 
suggest that the angular momentum content of the nuclear gas may be a critical 
factor that determines whether gas can flow to the center to feed an AGN.}

\end{enumerate}

\acknowledgments

This research is funded by the Carnegie Institution of Washington,, by 
NASA grants from the Space Telescope Science Institute (operated by AURA, 
Inc., under NASA contract NAS5-26555), and by NASA grant NAG 5-3556.  We thank 
Aaron Barth and Henrique Schmitt for helpful discussions on various aspects of 
this work.  We are grateful to an anonymous referee for constructive comments.

\clearpage
\appendix
 
\section{Supplementary H\al\ Luminosities for the Palomar Survey}

The catalog of H\al\ luminosities presented in Paper III contains a number of 
entries that were not reliable because they were based on observations taken 
under nonphotometric conditions\footnote{In Paper III, we listed the 
nonphotometric measurements as lower limits. This is not strictly correct. Line 
fluxes measured under nonphotometric conditions can be either too high or too 
low compared to the true value; they are simply inaccurate.}.  Table 4 gives 
an updated list of H\al\ luminosities for those galaxies for which we were 
able to locate published H\al\ fluxes.  Most of the data come from the surveys 
by  Heckman, Balick, \& Crane (1980), Stauffer (1982a), and Keel (1983a).  The 
luminosities are based on the distances given in Paper III, and they have been 
corrected for Galactic and internal extinction using the values of 
Galactic extinction and Balmer decrements given in Paper III.  In addition to 
the literature data, Table 4 also includes (3 $\sigma$) upper limits for the 
H\al\ luminosities of all absorption-line nuclei observed under photometric 
conditions.  The luminosity upper limits were compute by combining the 
continuum flux density at 6600 \AA\ with the 3 $\sigma$ upper limits for the 
equivalent width of H\al\ emission (see Paper III), assuming a line width of 
FWHM = 250 \kms.

\begin{figure*}[t]
\centerline{\psfig{file=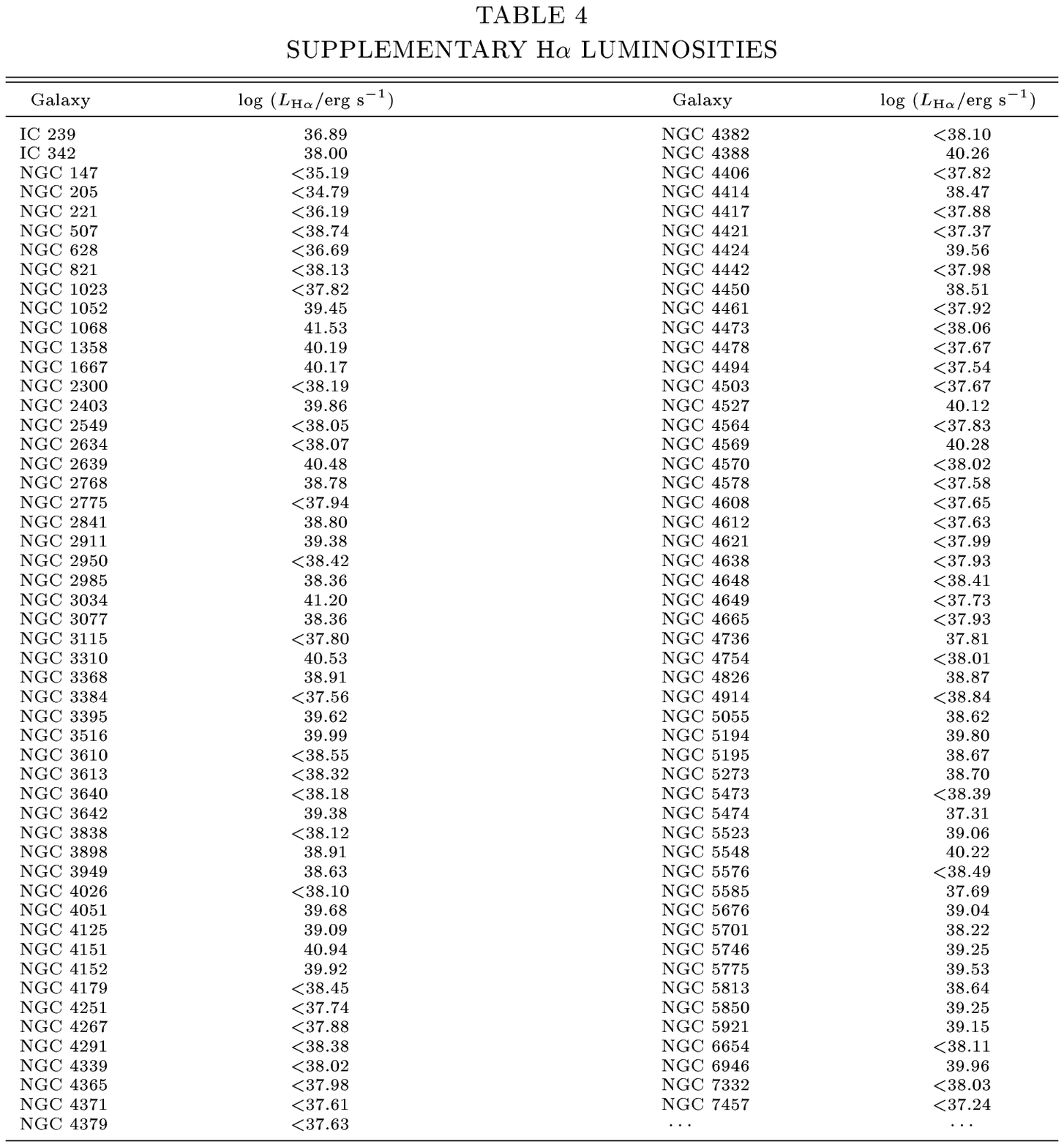,width=18.5cm,angle=0}}
\end{figure*}

\end{document}